\documentclass[aps,pre,twocolumn,groupedaddress,floatfix,longbibliography,superscriptaddress,10pt]{revtex4-2}

\usepackage{amsmath,amssymb,amsfonts,mathrsfs,bm,bbold,braket,slashed,cancel}

\usepackage{graphicx}
\usepackage{epsfig}
\usepackage{subfigure}
\usepackage{multirow}
\usepackage{feynmf}

\usepackage{times}

\usepackage[dvipsnames,svgnames]{xcolor}

\usepackage{hyperref}
\hypersetup{colorlinks=true,linkcolor=Blue,citecolor=Blue,urlcolor=Blue}

\usepackage{accents}
\usepackage{ulem}
\usepackage{simpler-wick}
\usepackage{orcidlink}

\DeclareUnicodeCharacter{039B}{\ensuremath{\Lambda}}

\newcommand{\bea}{\begin{eqnarray}}
\newcommand{\eea}{\end{eqnarray}}
\newcommand{\be}{\begin{equation}}
\newcommand{\ee}{\end{equation}}

\newcommand{\nn}{\nonumber}
\newcommand{\ii}{\mathrm{i}}

\newcommand{\A}{\mathcal{A}}

\newcommand{\sumint}{\mathop{\vcenter{\hbox{\ooalign{\hfil$\displaystyle\sum$\hfil\cr\hfil$\displaystyle\int$\hfil\cr}}}}}

\begin{document}

\title{Photon Quantization and Propagator in Rotating Coordinates}

\author{Alejandro Ayala~\orcidlink{0000-0003-3929-9209}}
\affiliation{Instituto de Ciencias Nucleares, Universidad Nacional Autónoma de México, Apartado Postal 70-543, CdMx 04510, México.}
\affiliation{Laboratório de Física Teórica e Computacional, Universidade Cidade de São Paulo, 01506-000, São Paulo, Brazil.}
\affiliation{Instituto de Física Teórica, Universidade Estadual Paulista, Rua Dr. Bento Teobaldo Ferraz, 271 - Bloco II, 01140-070 São Paulo, SP, Brazil.}
\author{Santiago Bernal-Langarica~\orcidlink{0000-0003-0826-9349}}
\affiliation{Instituto de Ciencias Nucleares, Universidad Nacional Autónoma de México, Apartado Postal 70-543, CdMx 04510, México.}
\author{Jorge David Casta\~no-Yepes~\orcidlink{0000-0002-8654-1304}}
\email{jorge.yepes@correounivalle.edu.co}
\affiliation{ Departamento de Física, Universidad del Valle, Ciudad Universitaria Meléndez,
Santiago de Cali 760032, Colombia}
\author{José Jorge Medina-Serna~\orcidlink{0009-0002-2026-1542}}
\affiliation{Instituto de Ciencias Nucleares, Universidad Nacional Autónoma de México, Apartado Postal 70-543, CdMx 04510, México.}

\begin{abstract}

We construct the quantized electromagnetic field and the corresponding photon propagators in a rigidly rotating cylindrical reference frame. Starting from the covariant Maxwell theory in rotating coordinates, we formulate the gauge-fixed field equations and analyze their solutions in cylindrical coordinates. The vector nature of the electromagnetic field requires distinguishing between orbital and intrinsic angular momentum, leading to a mode decomposition in which the total angular momentum provides the natural quantum number. We construct the two independent physical photon modes in terms of Bessel functions by solving the resulting coupled differential equations. We then provide an alternative construction of the physical photon modes based on a covariant basis of vector eigenfunctions generated by a covariantly constant tetrad in the rotating coordinates. We normalize them using the Maxwell inner product, and use them to quantize the electromagnetic field, defining the corresponding photon states and greater and lesser two-point functions. The physical time-ordered propagator is obtained directly from these quantized modes, with its transverse dependence expressed through polarization kernels that retain the cylindrical spatial structure and its spectral poles exhibiting the rotational energy shift associated with angular momentum. The full gauge-fixed photon propagator is constructed in two complementary ways. First, the scalar Green function in rotating coordinates is obtained and combined with the tensor transformation of the inertial Feynman-gauge propagator. Second, the same Green function is constructed from the covariant vector eigensystem using its spectral decomposition and the Fock--Schwinger proper-time representation. We verify that both constructions satisfy the Green-function equation associated with the complete gauge-fixed Maxwell operator, establishing the corresponding propagator for covariant perturbative calculations involving internal photon lines.

\end{abstract}


\maketitle

\section{Introduction}
Relativistic heavy-ion collisions provide a unique setting for investigating strongly interacting matter under extreme conditions~\cite{Busza:2018rrf,Braun-Munzinger:2015hba}. The high energy density reached in these reactions allows the properties of QCD matter to be explored over a broad range of temperatures and baryon densities, while the geometry of noncentral collisions introduces strong electromagnetic fields and a substantial orbital angular momentum~\cite{Busza:2018rrf,Braun-Munzinger:2015hba,Skokov:2009qp,Voronyuk:2011jd,Becattini:2007sr,Becattini:2015ska}. The evolution of the produced matter through these conditions can modify both its microscopic excitations and its collective properties, as illustrated by the in-medium behavior of hadronic resonances and other degrees of freedom~\cite{Rapp:1999ej,Ayala:2012ch,Braun-Munzinger:2015hba,Li:2016dta,Adhikari:2024}. Among the external fields generated in the collision, magnetic fields have received particular attention because of their potential impact on the dynamics of QCD matter. Strong magnetic fields are expected during the early stages of noncentral collisions, with their magnitude and space-time evolution determined by the collision geometry and the dynamics of the electrically charged matter~\cite{Skokov:2009qp,CastanoYepes2021,Voronyuk:2011jd,Adhikari:2024}. Their presence can affect chiral symmetry breaking and dynamical mass generation~\cite{Gusynin:1999pq,Rojas:2008sg,Mueller:2014tea}, while their influence on quark and fermion propagation has been investigated through modifications of masses, self-energies, and propagators~\cite{Ayala:2020dxs,Ayala:2020muk,Ayala:2021lor,Castano-Yepes:2022luw,Castano-Yepes:2023brq,Castano-Yepes:2024ltr}. The magnetic background also modifies the properties of fermion, meson, and gluon degrees of freedom~\cite{Avancini:2015ady,Avancini:2016fgq,Avancini:2018svs,Carlomagno:2022inu,Ayala:2019akk,Ayala:2018zat,Avancini:2021pmi,Coppola:2018vkw,Coppola:2023mmq,Li:2016dta,Moreira:2022dwo,Wang:2026xsm}, and can influence the electromagnetic response of the medium, including photon production and its angular dependence~\cite{Ayala:2017vex,Ayala:2019jey,Ayala:2022zhu,Ayala:2026vertex}. Magnetic fluctuations provide an additional source of modifications to the electromagnetic sector, including the generation of effective gauge-field masses~\cite{Castano-Yepes:2024ctr}, while the influence of magnetic backgrounds on the subsequent evolution of the medium has also been considered~\cite{Ayala:2024jvc,Sahu:2025tmb,Ayala:2025jpr}.

The orbital angular momentum carried by the colliding nuclei provides a different manifestation of the noncentral geometry. Part of this angular momentum can be transferred to the produced matter, giving rise to a vortical fluid and establishing a preferred direction for its collective motion~\cite{Becattini:2007sr,Becattini:2015ska,Becattini:2020ngo}. The resulting vorticity couples to the spin degrees of freedom of particles in the medium, providing a mechanism through which the rotational motion can be converted into measurable polarization~\cite{Liang:2004ph,Betz:2007kg,Gao:2007bc,Becattini:2013fla,Becattini:2016gvu}. Different theoretical descriptions have subsequently addressed the generation and evolution of polarization, including hydrodynamic, transport, and kinetic approaches~\cite{Pang:2016igs,Karpenko:2016jyx,Sun:2017xhx,Li:2017slc,Xia:2018tes,Wei:2018zfb,Fu:2020oxj,Florkowski:2018ahw,Liu:2021uhn}. This theoretical development has been accompanied by measurements of global and local polarization of hyperons over a broad range of collision energies by the STAR, ALICE, and HADES Collaborations~\cite{STAR:2017ckg,STAR:2018gyt,STAR:2021beb,STAR:2023nvo,ALICE:2019onw,HADES:2022enx}. The magnitude of the global $\Lambda$ polarization measured at RHIC has been used to infer a characteristic angular velocity of order $9\times10^{21}\,\mathrm{s}^{-1}$~\cite{STAR:2017ckg}. Measurements of polarization along the beam direction and of vector-meson spin alignment provide complementary information on the local rotational structure and spin response of the produced matter~\cite{STAR:2019erd,STAR:2020xbm,STAR:2022fan,STAR:2023eck,Sarkar:2022uyt}, while recent analyses have further explored the energy and spatial dependence of the polarization signal~\cite{Ayala:2020soy,Ayala:2021xrn,Ayala:2023xyn,Ayala:2026dmz}.

The existence of a macroscopic rotational motion raises a more general question concerning the description of microscopic degrees of freedom in the rotating medium. Once vorticity becomes an intrinsic property of the state, the medium defines a preferred local rest frame and a distinguished spatial direction, and correlation functions can acquire structures associated with the rotational background. This has motivated the study of quantum fields in rotating systems, beginning with the quantization of scalar fields in rotating coordinates~\cite{LetawPfautsch1980} and the analysis of quantum fields at finite temperature in a rotating system~\cite{Vilenkin1980}. The structure of the quantum vacuum and of thermal states in rotating systems was subsequently investigated in greater detail~\cite{DaviesDrayManogue1996,AmbrusWinstanley2014}, while rotation has also been incorporated into fermionic propagation and interaction effects in relativistic thermal media~\cite{Ayala:2021osy,Gaspar:2023nqk}. More recent formulations have considered rigid rotation in cylindrical geometries for both scalar and fermionic fields, with applications to propagators, thermal observables, and perturbative corrections in rotating quantum systems~\cite{vilenkin1980rotation,siri2024bose,kawaguchi2025susceptibilities,salvio2026scalar,kuboniwa2026perturbation,hernandez2026scalar,CastanoYepes:2025,CastanoYepes2026FiniteSize}. In these descriptions, the angular momentum about the rotation axis provides a natural quantum number for organizing the modes, while the radial structure of the system determines the corresponding transverse spectrum. For finite cylindrical systems, this structure leads to Fourier--Bessel representations in which the spatial dependence of the modes and propagators remains explicit~\cite{kuboniwa2026perturbation,hernandez2026scalar}. These developments establish cylindrical quantization as a useful framework for describing microscopic fields in rigidly rotating media.

The cylindrical formulation is directly connected with the causal structure of rigid rotation, since the tangential velocity reaches the speed of light at the corresponding light cylinder, constraining the relation between the angular velocity and the transverse size of a finite rotating system~\cite{vilenkin1980rotation,siri2024bose,kuboniwa2026perturbation,hernandez2026scalar}. At the same time, the radial extent determines the allowed transverse spectrum and can therefore affect the microscopic and thermal properties of the system, with this interplay between rotation, finite volume, and the causal boundary also studied in effective descriptions of rotating QCD matter~\cite{PhysRevD.110.094053,4zrn-wgrg,knn8-sv3k}. These effects extend to thermal observables, where rotation modifies thermodynamic quantities and characteristic temperatures associated with chiral and deconfinement transitions, while finite-volume effects and thermal fluctuations can further alter the structure of thermal observables and phase transitions~\cite{gaspar2023chiral,hernandez2025vortical,siri2024bose,PhysRevD.103.094515,PhysRevD.110.094053,knn8-sv3k,PhysRevD.106.116019,PhysRevD.110.056014}. Scalar field theories have likewise provided a framework for studying the interplay between rotation, finite size, and thermal fluctuations through propagators and perturbative corrections in finite rotating cylinders~\cite{ambrus2023rigid,kuboniwa2026perturbation,hernandez2026scalar,CastanoYepes2026FiniteSize}.
This analysis is particularly relevant for electromagnetic probes. Photons are produced throughout the different stages of a heavy-ion collision and, once produced, do not undergo strong interactions with the surrounding medium. Their electromagnetic coupling, therefore, allows them to carry information about the conditions of the system at the time of emission, complementing hadronic observables that are sensitive to the subsequent evolution. In addition to the effects associated with magnetic backgrounds, electromagnetic observables can respond directly to the vortical motion of the medium. Vortical contributions to electromagnetic currents and the associated photon degrees of freedom have been investigated in quantum and kinetic descriptions~\cite{ChernodubCortijoLandsteiner2018,HuangMitkinSadofyevSperanza2020}, while calculations of electromagnetic observables in rotating thermal media have considered the consequences of the rotational background for photon and dilepton production~\cite{CastanoYepes:2025}. These results make the electromagnetic field a natural sector in which the microscopic consequences of rotational motion can be investigated.

From this perspective, the description of photon propagation in a rotating medium requires a formulation that retains both the cylindrical geometry and the angular-momentum structure of the electromagnetic field. In particular, the vector nature of the photon field introduces an intrinsic angular-momentum contribution in addition to the orbital angular momentum associated with the cylindrical modes. The construction of the corresponding photon modes and propagator must therefore account simultaneously for the rotational geometry, the physical polarization structure, and the modified relation between the rotating-frame energy and angular momentum. A systematic formulation of these ingredients provides the basis for studying electromagnetic correlation functions and photon-mediated processes in rotating matter.

For the electromagnetic field, however, rotation requires a formulation adapted to a noninertial reference frame. Maxwell's equations in rotating coordinates have been considered from classical and covariant viewpoints~\cite{Georgiou1988,Obukhov2021}, and the description of electromagnetic fields in cylindrical rotating systems has also been connected with vortical electromagnetic observables~\cite{ChernodubCortijoLandsteiner2018,HuangMitkinSadofyevSperanza2020}. In such a description, the rotating coordinates determine a nontrivial metric even though the underlying spacetime remains flat, and the relation between coordinate and locally inertial components of the electromagnetic field is encoded in the corresponding tetrad. The construction of physical photon modes must therefore be consistent with both the covariant Maxwell equation and the cylindrical symmetry associated with rigid rotation. This provides a direct route from the geometry of the rotating frame to the quantum description of the photon field.

Inspired by the latter discussion, the present work studies the solutions of Maxwell's equations in a rigidly rotating background and identifies the physical photon modes that arise from the rotational geometry. We further construct the corresponding photon propagator from the quantized physical modes and obtain the full gauge-fixed propagator in rotating coordinates. 

This manuscript is organized as follows. In Sec.~\ref{sec:geometry_and_verbein}, we introduce the rigidly rotating coordinates and derive the corresponding spacetime geometry, including the metric, inverse metric, and tetrad. In Sec.~\ref{subsec:PhotonWaveEquation}, we formulate the covariant Maxwell equations in the rotating frame and derive the corresponding photon wave equation in cylindrical coordinates, emphasizing the role of the vector covariant Laplacian and the decomposition into orbital and intrinsic angular momentum. In Sec.~\ref{sec:solutionsofwaveequation}, we construct the solutions of the wave equation and analyze their general structure in terms of the total angular momentum as the appropriate quantum number. In Sec.~\ref{subsec:PhotonModes}, we identify the physical photon modes, including the two independent polarization sectors and their angular-momentum structure, and provide an alternative construction based on a covariant basis of vector eigenfunctions generated by the covariantly constant tetrad. In Sec.~\ref{subsec:PhotonModeNormalization}, we normalize these modes using the Maxwell inner product and the orthogonality properties of the cylindrical Bessel functions. In Sec.~\ref{subsec:PhotonFieldExpansion}, we use the normalized physical modes to expand and quantize the electromagnetic field, defining the corresponding creation and annihilation operators and photon states. In Sec.~\ref{subsec:PhotonPropagator}, we construct the physical Feynman photon propagator in the rotating frame directly from the quantized field, including its greater and lesser two-point functions and the associated polarization kernels. In Sec.~\ref{sec:FullGaugeFixedPhotonPropagator}, we construct the full gauge-fixed photon propagator in two ways. First, we obtain the scalar Green function in rotating coordinates and combine it with the tensor transformation of the inertial Feynman-gauge propagator. Second, we construct the corresponding Green function from the covariant vector eigenbasis using its spectral decomposition and the Fock--Schwinger proper-time representation. We verify that the resulting propagator satisfies the Green-function equation associated with the complete gauge-fixed Maxwell operator. Finally, Sec.~\ref{sec:conclusiones} summarizes the main results and discusses their possible applications.

\section{Rotating-Frame Geometry and Tetrad}\label{sec:geometry_and_verbein}
To describe a rigidly rotating frame, we start from inertial Minkowski coordinates $(t',x',y',z')$ and consider a rotation about the $z$ axis with constant angular velocity $\Omega$. The transformation from the rotating coordinates $(t,x,y,z)$ to the inertial coordinates is given by
\bea
t'&=&t,\nn\\
x'&=&x\cos(\Omega t)-y\sin(\Omega t),\nn\\
y'&=&x\sin(\Omega t)+y\cos(\Omega t),\nn\\
z'&=&z.
\label{PhotonCoordinateTransformation}
\eea

Differentiating Eq.~\eqref{PhotonCoordinateTransformation} gives
\bea
dx'&=&\cos(\Omega t)\,dx+\sin(\Omega t)\,dy\nn\\
&&+\Omega\left[-x\sin(\Omega t)+y\cos(\Omega t)\right]dt,\nn\\
dy'&=&-\sin(\Omega t)\,dx+\cos(\Omega t)\,dy\nn\\
&&-\Omega\left[x\cos(\Omega t)+y\sin(\Omega t)\right]dt,\nn\\
dz'&=&dz.
\label{PhotonDifferentialTransformation}
\eea

Substituting these relations into the Minkowski line element,
\bea
ds^2=dt'^2-dx'^2-dy'^2-dz'^2,
\eea
and using $dt'=dt$, one obtains
\bea
ds^2&=&\left[1-\Omega^2(x^2+y^2)\right]dt^2+2\Omega y\,dx\,dt-2\Omega x\,dy\,dt\nn\\
&&-dx^2-dy^2-dz^2.
\label{RotatingPhotonLineElement}
\eea

The corresponding covariant metric tensor is therefore
\bea
g_{\mu\nu}=
\begin{pmatrix}
1-\Omega^2(x^2+y^2)&\Omega y&-\Omega x&0\\
\Omega y&-1&0&0\\
-\Omega x&0&-1&0\\
0&0&0&-1
\end{pmatrix}.
\label{PhotonRotatingMetric}
\eea
with determinant
\bea
g\equiv\det(g_{\mu\nu})=-1,
\label{PhotonMetricDeterminant}
\eea
and hence the corresponding volume factor is simply
\bea
\sqrt{-g}=1.
\label{PhotonMetricDeterminantRoot}
\eea

The inverse metric is defined by
\bea
g^{\mu\alpha}g_{\alpha\nu}=\delta^\mu_{\ \nu}.
\label{PhotonInverseMetricDefinition}
\eea

For the metric in Eq.~\eqref{PhotonRotatingMetric}, this gives
\bea
g^{\mu\nu}=
\begin{pmatrix}
1&\Omega y&-\Omega x&0\\
\Omega y&-1+\Omega^2y^2&-\Omega^2xy&0\\
-\Omega x&-\Omega^2xy&-1+\Omega^2x^2&0\\
0&0&0&-1
\end{pmatrix}.
\label{PhotonInverseMetric}
\eea

Although the metric is non-diagonal and depends explicitly on the transverse coordinates, this coordinate dependence does not imply a nonvanishing spacetime curvature. Since the rotating coordinates are related to inertial Minkowski coordinates by an exact coordinate transformation, the spacetime remains flat, and the curvature tensors satisfy
\bea
R^\mu_{\ \nu\alpha\beta}=0,\qquad R_{\mu\nu}=0,\qquad R=0.
\label{PhotonFlatRotatingSpacetime}
\eea
The nontrivial components of the metric therefore encode the inertial effects associated with the non-inertial rotating frame rather than those of a gravitational field.

The temporal component of the metric is
\bea
g_{00}=1-\Omega^2(x^2+y^2).
\label{PhotonMetricTemporalComponent}
\eea
Introducing the cylindrical radial coordinate
\bea
\rho=\sqrt{x^2+y^2},
\label{PhotonCylindricalRadius}
\eea
it becomes
\bea
g_{00}=1-\Omega^2\rho^2.
\label{PhotonMetricTemporalCylindrical}
\eea

To identify the light cone associated with the rotating frame, we consider observers at fixed spatial coordinates, for which $dx=dy=dz=0$. Their spacetime interval is therefore
\bea
ds^2=g_{00}\,dt^2=\left(1-\Omega^2\rho^2\right)dt^2.
\label{PhotonRotatingObserverInterval}
\eea
The worldline becomes null when $g_{00}=0$, which gives
\bea
\Omega^2\rho^2=1.
\label{PhotonLightCylinder}
\eea
and hence the corresponding radial position is
\bea
\rho_{\rm LC}=\frac{1}{\Omega}.
\label{PhotonLightCylinderRadius}
\eea

An observer co-rotating with the frame has tangential velocity $v=\Omega\rho$, so that at $\rho=\rho_{\rm LC}$ one obtains $v=1$ in units where $c=1$. Thus, the surface $g_{00}=0$ corresponds to the light cylinder. For a rigidly rotating system confined within a radius $R$, requiring the system to remain inside this surface gives
\bea
\Omega R<1.
\label{PhotonCausalityCondition}
\eea

It is useful to introduce a tetrad adapted to the rotating coordinates. The tetrad relates the spacetime metric to the Minkowski metric in the local Lorentz frame according to
\bea
g^{\mu\nu}=e^\mu_{\ a}e^\nu_{\ b}\eta^{ab},
\label{PhotonMetricFromTetrad}
\eea
where
\bea
\eta^{ab}=\mathrm{diag}(1,-1,-1,-1).
\label{PhotonTangentMetric}
\eea

The tetrad is not uniquely determined by the spacetime metric. The tetrad is required to satisfy $g^{\mu\nu}=e^\mu_{\ a}e^\nu_{\ b}\eta^{ab}$, but this relation is invariant under local Lorentz transformations $e^\mu_{\ a}\rightarrow e^\mu_{\ b}\Lambda^b_{\ a}$, with $\Lambda^c_{\ a}\Lambda^d_{\ b}\eta_{cd}=\eta_{ab}$. Hence, infinitely many tetrads are compatible with the same metric, and one may choose the local Lorentz frame that is most convenient for the problem at hand. We determine a convenient choice directly from the components of the inverse metric in Eq.~\eqref{PhotonInverseMetric}. For the temporal component,
\bea
g^{00}=(e^0_{\ 0})^2-(e^0_{\ 1})^2-(e^0_{\ 2})^2-(e^0_{\ 3})^2.
\label{PhotonTetrad00}
\eea

Since $g^{00}=1$, we choose the local Lorentz frame such that
\bea
e^0_{\ 0}=1,\qquad e^0_{\ 1}=e^0_{\ 2}=e^0_{\ 3}=0.
\label{PhotonTetradTemporalComponents}
\eea
so that mixed temporal-spatial components then become
\bea
g^{01}=e^1_{\ 0},\qquad g^{02}=e^2_{\ 0}.
\label{PhotonTetradMixedComponents}
\eea

Using $g^{01}=\Omega y$ and $g^{02}=-\Omega x$, we obtain
\bea
e^1_{\ 0}=\Omega y,\qquad e^2_{\ 0}=-\Omega x.
\label{PhotonTetradMixedComponentsResult}
\eea

For the spatial directions, we choose the unit diagonal components
\bea
e^1_{\ 1}=e^2_{\ 2}=e^3_{\ 3}=1,
\label{PhotonTetradSpatialDiagonal}
\eea
with the remaining spatial components set to zero. The resulting components reproduce the spatial part of the inverse metric,
\bea
g^{11}&=&(e^1_{\ 0})^2-(e^1_{\ 1})^2=\Omega^2y^2-1,\nn\\
g^{22}&=&(e^2_{\ 0})^2-(e^2_{\ 2})^2=\Omega^2x^2-1,
\label{PhotonTetradDiagonalComponents}
\eea
and
\bea
g^{12}=e^1_{\ 0}e^2_{\ 0}=-\Omega^2xy.
\label{PhotonTetradOffDiagonalComponent}
\eea

Thus, the tetrad is
\bea
e^\mu_{\ a}=
\begin{pmatrix}
1&0&0&0\\
\Omega y&1&0&0\\
-\Omega x&0&1&0\\
0&0&0&1
\end{pmatrix}.
\label{PhotonTetrad}
\eea
Its inverse is defined by
\bea
e^\mu_{\ a}e^a_{\ \nu}=\delta^\mu_{\ \nu},\qquad e^a_{\ \mu}e^\mu_{\ b}=\delta^a_{\ b},
\label{PhotonInverseTetradDefinition}
\eea
and is given by
\bea
e^a_{\ \mu}=
\begin{pmatrix}
1&0&0&0\\
-\Omega y&1&0&0\\
\Omega x&0&1&0\\
0&0&0&1
\end{pmatrix}.
\label{PhotonInverseTetrad}
\eea
It satisfies the equivalent relation
\bea
g_{\mu\nu}=\eta_{ab}e^a_{\ \mu}e^b_{\ \nu}.
\label{PhotonMetricFromInverseTetrad}
\eea

\section{The photon rotating wave equation}\label{subsec:PhotonWaveEquation}
The electromagnetic field in the rotating frame is described by the vector potential $A_\mu$. Although the rotating coordinates introduce
nontrivial metric components and a nonvanishing Levi-Civita connection,
the electromagnetic field strength must remain a tensor quantity.
This requires the derivative of $A_\mu$ to be formulated covariantly. For a covariant vector, the covariant derivative is
\bea
\nabla_\mu A_\nu=\partial_\mu A_\nu-\Gamma^\lambda_{\mu\nu}A_\lambda,
\label{PhotonCovariantDerivativeCovector}
\eea
whereas for a contravariant vector, the corresponding covariant derivative is
\bea
\nabla_\mu A^\nu=\partial_\mu A^\nu+\Gamma^\nu_{\mu\lambda}A^\lambda.
\label{PhotonCovariantDerivativeVector}
\eea

The connection coefficients $\Gamma^\lambda_{\mu\nu}$ are determined entirely
by the spacetime metric and, for the torsion-free metric-compatible
Levi-Civita connection, are given by
\bea
\Gamma^\lambda_{\mu\nu}=\frac{1}{2}g^{\lambda\sigma}\left(\partial_\mu g_{\sigma\nu}+\partial_\nu g_{\sigma\mu}-\partial_\sigma g_{\mu\nu}\right).
\label{PhotonChristoffelSymbols}
\eea
They are symmetric in their lower indices,
\bea
\Gamma^\lambda_{\mu\nu}=\Gamma^\lambda_{\nu\mu},
\label{PhotonChristoffelSymmetry}
\eea
and are not tensor components themselves. Their role is to compensate for
the coordinate dependence of the basis vectors and to ensure that the
covariant derivative transforms tensorially.

The curvature associated with the connection is described by the Riemann
tensor,
\bea
R^\rho{}_{\sigma\mu\nu}=\partial_\mu\Gamma^\rho_{\nu\sigma}-\partial_\nu\Gamma^\rho_{\mu\sigma}+\Gamma^\rho_{\mu\lambda}\Gamma^\lambda_{\nu\sigma}-\Gamma^\rho_{\nu\lambda}\Gamma^\lambda_{\mu\sigma}.
\label{PhotonRiemannTensor}
\eea
The Ricci tensor is obtained by contracting the first and third indices of
the Riemann tensor,
\bea
R_{\sigma\nu}=R^\rho{}_{\sigma\rho\nu},
\label{PhotonRicciTensor}
\eea
while the corresponding Ricci scalar is
\bea
R=g^{\mu\nu}R_{\mu\nu}.
\label{PhotonRicciScalar}
\eea

For the rotating coordinates considered here, the spacetime is still
Minkowski spacetime and the rotation is introduced solely through the
coordinate transformation. Then, the nonvanishing connection
coefficients do not imply a nonzero spacetime curvature. The Riemann tensor,
Ricci tensor, and Ricci scalar therefore satisfy
\bea
R^\rho{}_{\sigma\mu\nu}=0,\qquad R_{\mu\nu}=0,\qquad R=0.
\label{PhotonFlatSpacetime}
\eea
Thus, the rotating coordinate system is non-inertial but remains a
coordinate description of flat spacetime. The connection coefficients enter
the covariant derivatives and the explicit form of the wave operator, while
the curvature tensors vanish identically.

The field strength is constructed from the antisymmetric part of the
derivative of the potential. Applying this antisymmetrization to
Eq.~\eqref{PhotonCovariantDerivativeCovector} gives
\bea
\nabla_\mu A_\nu-\nabla_\nu A_\mu
&=&\partial_\mu A_\nu-\partial_\nu A_\mu-\left(\Gamma^\lambda_{\mu\nu}
-\Gamma^\lambda_{\nu\mu}\right)A_\lambda,\nn\\
\label{PhotonAntisymmetricCovariantDerivative}
\eea
so that after appliying Eq.~\eqref{PhotonChristoffelSymmetry}, the connection drops out for the antisymmetric derivative,
\bea
\nabla_\mu A_\nu-\nabla_\nu A_\mu
=\partial_\mu A_\nu-\partial_\nu A_\mu.
\label{PhotonAntisymmetricDerivativeReduction}
\eea

Hence, the electromagnetic field-strength tensor can be expressed
equivalently in covariant or coordinate form as
\bea
F_{\mu\nu}
\equiv\nabla_\mu A_\nu-\nabla_\nu A_\mu
=\partial_\mu A_\nu-\partial_\nu A_\mu.
\label{PhotonCovariantFieldStrength}
\eea

The covariant action for the electromagnetic field is constructed from the
scalar $F_{\mu\nu}F^{\mu\nu}$, together with the invariant volume element
$\sqrt{-g}\,d^4x$. It is given by
\bea
S_{\rm EM}
&=&-\frac14\int d^4x\,\sqrt{-g}\,
F_{\mu\nu}F^{\mu\nu}
\nn\\
&=&-\frac14\int d^4x\,\sqrt{-g}\,
g^{\mu\alpha}g^{\nu\beta}
F_{\mu\nu}F_{\alpha\beta}.
\label{PhotonElectromagneticAction}
\eea
For the rotating metric considered here, $\sqrt{-g}=1$, so the invariant
volume element reduces to $d^4x$,
\bea
S_{\rm EM}
=-\frac14\int d^4x\,F_{\mu\nu}F^{\mu\nu},
\label{PhotonElectromagneticActionRotating}
\eea
and the metric dependence, however, remains present in the contraction of the
field-strength tensor through
\bea
F^{\mu\nu}
=g^{\mu\alpha}g^{\nu\beta}F_{\alpha\beta}.
\label{PhotonRaisedFieldStrength}
\eea

The equations of motion follow from varying the action with respect to
the potential $A_\mu$. Since the field strength is expressed in terms of
the antisymmetric covariant derivative, its variation is
\bea
\delta F_{\mu\nu}
=\nabla_\mu\delta A_\nu-\nabla_\nu\delta A_\mu.
\label{PhotonVariationFieldStrength}
\eea

The variation of the action is therefore
\bea
\delta S_{\rm EM}
&=&-\frac12\int d^4x\,\sqrt{-g}\,
F^{\mu\nu}\delta F_{\mu\nu}
\nn\\
&=&-\frac12\int d^4x\,\sqrt{-g}\,F^{\mu\nu}
\left(\nabla_\mu\delta A_\nu-\nabla_\nu\delta A_\mu\right).\nn\\
\label{PhotonActionVariationExpanded}
\eea

Using the antisymmetry $F^{\mu\nu}=-F^{\nu\mu}$ the two terms in the variation give identical contributions after
relabeling the dummy indices. Hence,
\bea
\delta S_{\rm EM}=-\int d^4x\,\sqrt{-g}\,F^{\mu\nu}\nabla_\mu\delta A_\nu.
\label{PhotonActionVariationSimplified}
\eea

Applying covariant integration by parts gives
\bea
\int d^4x\,\sqrt{-g}\,
F^{\mu\nu}\nabla_\mu\delta A_\nu
&=&\int d^4x\,\sqrt{-g}\,\nabla_\mu\left(F^{\mu\nu}\delta A_\nu\right)\nn\\
&-&\int d^4x\,\sqrt{-g}\,
(\nabla_\mu F^{\mu\nu})\delta A_\nu.\nn\\
\label{PhotonCovariantIntegrationByParts}
\eea

For variations that vanish at the boundary, the first term does not
contribute, yielding
\bea
\delta S_{\rm EM}=\int d^4x\,\sqrt{-g}\,(\nabla_\mu F^{\mu\nu})\delta A_\nu.
\label{PhotonActionVariationFinal}
\eea

Since $\delta A_\nu$ is arbitrary, the Euler-Lagrange equation is
\bea
\nabla_\mu F^{\mu\nu}=0.
\label{PhotonMaxwellEquation}
\eea

Using the antisymmetry of $F^{\mu\nu}$, this equation can equivalently be
written as
\bea
\nabla_\nu F^{\mu\nu}=0.
\label{PhotonMaxwellEquationLowered}
\eea

To make the Maxwell equation explicit in the rotating coordinates, we
expand its covariant divergence. For a rank-two tensor, the divergence
with respect to the second index is
\bea
\nabla_\nu F^{\mu\nu}
&=&\partial_\nu F^{\mu\nu}+\Gamma^\mu_{\nu\lambda}F^{\lambda\nu}+\Gamma^\nu_{\nu\lambda}F^{\mu\lambda}.
\label{PhotonTensorDivergence}
\eea
For the Levi-Civita connection, the contracted Christoffel symbol can be
expressed in terms of the determinant of the metric as
\bea
\Gamma^\nu_{\nu\lambda}=\partial_\lambda\ln\sqrt{-g}.
\label{PhotonContractedChristoffel}
\eea
Since the rotating metric satisfies $\sqrt{-g}=1$, this contribution
vanishes identically. The Maxwell equation therefore reduces to
\bea
\nabla_\nu F^{\mu\nu}=\partial_\nu F^{\mu\nu}+\Gamma^\mu_{\nu\lambda}F^{\lambda\nu}=0.
\label{PhotonMaxwellExplicit}
\eea

In order to construct the gauge-field propagator, the gauge freedom of the
electromagnetic field must be fixed, since the gauge redundancy makes the
quadratic differential operator non-invertible. We choose the covariant
gauge-fixing action
\bea
S_{\rm gf}=-\frac{1}{2\xi}\int d^4x\,\sqrt{-g}\,\left(\nabla_\mu A^\mu\right)^2.
\label{PhotonGaugeFixingAction}
\eea
where $\xi$ denotes the gauge-fixing parameter. The total action is then
\bea
S=S_{\rm EM}+S_{\rm gf}.
\label{PhotonTotalGaugeFixedAction}
\eea

Varying the gauge-fixing contribution with respect to the gauge potential
gives
\bea
\delta S_{\rm gf}=\frac{1}{\xi}\int d^4x\,\sqrt{-g}\,(\nabla_\mu A^\mu)\nabla_\nu\delta A^\nu.
\label{PhotonGaugeFixingVariation}
\eea
Using covariant integration by parts and assuming that the variation
vanishes at the boundary, we obtain
\bea
\delta S_{\rm gf}=-\frac{1}{\xi}\int d^4x\,\sqrt{-g}\,\nabla_\nu(\nabla_\mu A^\mu)\delta A^\nu.
\label{PhotonGaugeFixingVariationIntegrated}
\eea

Combining this contribution with the variation of the electromagnetic
action in Eq.~\eqref{PhotonActionVariationFinal}, the gauge-fixed equation
of motion becomes
\bea
\nabla_\nu F^{\mu\nu}-\frac{1}{\xi}\nabla^\mu\left(\nabla_\alpha A^\alpha\right)=0.
\label{PhotonGaugeFixedMaxwell}
\eea

The field-strength tensor can be expressed in terms of the vector
potential by raising the indices in Eq.~\eqref{PhotonCovariantFieldStrength}
with the inverse metric. Thus,
\bea
F^{\mu\nu}
&=&g^{\mu\alpha}g^{\nu\beta}F_{\alpha\beta}
=g^{\mu\alpha}g^{\nu\beta}
\left(\nabla_\alpha A_\beta-\nabla_\beta A_\alpha\right).\nn\\
\label{PhotonFieldStrengthIndexRaising}
\eea
Using metric compatibility, $\nabla_\lambda g_{\mu\nu}=0$, together with
$A^\nu=g^{\nu\beta}A_\beta$ and
$\nabla^\mu=g^{\mu\alpha}\nabla_\alpha$, we obtain
\bea
g^{\mu\alpha}g^{\nu\beta}\nabla_\alpha A_\beta&=&\nabla^\mu A^\nu,\nn\\
g^{\mu\alpha}g^{\nu\beta}\nabla_\beta A_\alpha&=&\nabla^\nu A^\mu,
\label{PhotonRaisedFieldStrengthTerms}
\eea
and the field-strength tensor can be written as
\bea
F^{\mu\nu}=\nabla^\mu A^\nu-\nabla^\nu A^\mu.
\label{PhotonFieldStrengthContravariant}
\eea

Substituting Eq.~\eqref{PhotonFieldStrengthContravariant} into
Eq.~\eqref{PhotonGaugeFixedMaxwell}, we find
\bea
\nabla_\nu F^{\mu\nu}=\nabla_\nu\nabla^\mu A^\nu-\nabla_\nu\nabla^\nu A^\mu.
\label{PhotonMaxwellVectorExpansion}
\eea

The ordering of the covariant derivatives in the first term can be
changed using their commutator. For a vector field,
\bea
[\nabla_\nu,\nabla^\mu]A^\nu=R^\mu_{\ \alpha}A^\alpha,
\label{PhotonCovariantDerivativeCommutator}
\eea
where $R^\mu_{\ \alpha}$ is defined in Eq.~\eqref{PhotonRicciTensor}.

Hence,
\bea
\nabla_\nu\nabla^\mu A^\nu=\nabla^\mu\nabla_\nu A^\nu+R^\mu_{\ \alpha}A^\alpha.
\label{PhotonCommutedDerivative}
\eea
Substituting Eq.~\eqref{PhotonCommutedDerivative} into
Eq.~\eqref{PhotonGaugeFixedMaxwell} gives
\bea
-\nabla_\beta\nabla^\beta A^\mu+\nabla^\mu\nabla_\alpha A^\alpha+R^\mu_{\ \alpha}A^\alpha-\frac{1}{\xi}\nabla^\mu\nabla_\alpha A^\alpha=0.\nn\\
\label{PhotonGaugeFixedExpanded}
\eea

Collecting the terms containing $\nabla^\mu\nabla_\alpha A^\alpha$, we
obtain
\bea
\left[\delta^\mu_{\ \alpha}\nabla_\beta\nabla^\beta-R^\mu_{\ \alpha}+\left(\frac{1}{\xi}-1\right)\nabla^\mu\nabla_\alpha\right]A^\alpha=0.
\label{PhotonGaugeFixedOperator}
\eea

For the rotating coordinates considered here, the metric is related to
the Minkowski metric by a coordinate transformation. The spacetime is
therefore flat, so that
\bea
R^\mu_{\ \alpha}=0,
\label{PhotonRicciZero}
\eea
and the gauge-fixed equation Then becomes
\bea
\left[\delta^\mu_{\ \alpha}\nabla_\beta\nabla^\beta+\left(\frac{1}{\xi}-1\right)\nabla^\mu\nabla_\alpha\right]A^\alpha=0.
\label{PhotonGaugeFixedFlatOperator}
\eea

The choice of the Feynman gauge $\xi=1$ eliminates the second derivative term and yields
\bea
\nabla_\beta\nabla^\beta A^\mu=0.
\label{PhotonFeynmanGaugeWaveEquation}
\eea

We next determine the explicit structure of the wave operator in the
rotating coordinates. The covariant d'Alembertian is defined as
\bea
\Box\equiv g^{\mu\nu}\nabla_\mu\nabla_\nu=\frac{1}{\sqrt{-g}}\partial_\mu\left(\sqrt{-g}\,g^{\mu\nu}\partial_\nu\right).
\label{PhotonScalarDAlembertianDefinition}
\eea
Since $\sqrt{-g}=1$ for the rotating metric, Eq.~\eqref{PhotonScalarDAlembertianDefinition} reduces to
\bea
\Box=\partial_\mu\left(g^{\mu\nu}\partial_\nu\right)=g^{\mu\nu}\partial_\mu\partial_\nu+\left(\partial_\mu g^{\mu\nu}\right)\partial_\nu.
\label{PhotonScalarDAlembertianUnitDeterminant}
\eea
Using the inverse metric given in Eq.~\eqref{PhotonInverseMetric}, the second-derivative
part of the operator is
\bea
g^{\mu\nu}\partial_\mu\partial_\nu&=&\partial_t^2+2\Omega y\,\partial_t\partial_x-2\Omega x\,\partial_t\partial_y+\Omega^2y^2\partial_x^2\nn\\
&+&\Omega^2x^2\partial_y^2-2\Omega^2xy\,\partial_x\partial_y-\partial_x^2-\partial_y^2-\partial_z^2.\nn\\
\label{PhotonExpandedDAlembertian}
\eea
The coordinate dependence of the inverse metric also produces
first-derivative terms. In the present case,
\bea
\partial_\mu g^{\mu x}&=&-\Omega^2x,\nn\\
\partial_\mu g^{\mu y}&=&-\Omega^2y,\nn\\
\partial_\mu g^{\mu t}&=&\partial_\mu g^{\mu z}=0.
\label{PhotonFirstDerivativeTerms}
\eea
Therefore, the complete scalar d'Alembertian is
\bea
\Box&=&\partial_t^2+2\Omega y\,\partial_t\partial_x-2\Omega x\,\partial_t\partial_y+\Omega^2y^2\partial_x^2+\Omega^2x^2\partial_y^2\nn\\
&-&2\Omega^2xy\,\partial_x\partial_y-\Omega^2x\partial_x-\Omega^2y\partial_y-\partial_x^2-\partial_y^2-\partial_z^2.\nn\\
\label{PhotonCompleteDAlembertian}
\eea

The terms involving the transverse spatial derivatives can be expressed
in terms of the orbital angular momentum operator around the rotation
axis. We define
\bea
\hat L_z\equiv-\ii(x\partial_y-y\partial_x).
\label{PhotonOrbitalAngularMomentum}
\eea
so that the square of the orbital operator is
\bea
\hat L_z^2&=&-\left(x\partial_y-y\partial_x\right)^2\nn\\
&=&x^2\partial_y^2+y^2\partial_x^2-2xy\,\partial_x\partial_y-x\partial_x-y\partial_y
\label{PhotonOrbitalAngularMomentumSquare}
\eea

Then,
\bea
-\Omega^2\hat L_z^2&=&\Omega^2x^2\partial_y^2+\Omega^2y^2\partial_x^2-2\Omega^2xy\,\partial_x\partial_y-\Omega^2x\partial_x\nn\\
&-&\Omega^2y\partial_y.
\label{PhotonOrbitalSquareContribution}
\eea

Using Eqs.~\eqref{PhotonOrbitalAngularMomentum} and
\eqref{PhotonOrbitalSquareContribution}, the d'Alembertian can be written
in the compact form
\bea
\Box=\partial_t^2-2\ii\Omega\hat L_z\partial_t-\Omega^2\hat L_z^2-\partial_x^2-\partial_y^2-\partial_z^2.
\label{PhotonCompactDAlembertian}
\eea
Equivalently, introducing the transverse Laplacian
$\nabla_\perp^2=\partial_x^2+\partial_y^2$, one may write
\bea
\Box=\left(\partial_t-\ii\Omega\hat L_z\right)^2-\nabla_\perp^2-\partial_z^2.
\label{PhotonRotatingDAlembertian}
\eea

In cylindrical coordinates,
\bea
x&=&\rho\cos\phi,\qquad y=\rho\sin\phi,\qquad z=z.
\label{PhotonCylindricalCoordinates}
\eea
the orbital angular momentum operator becomes
\bea
\hat L_z=-\ii\partial_\phi.
\label{PhotonOrbitalGeneratorCylindrical}
\eea
and the scalar d'Alembertian can be written as
\bea
\Box=\left(\partial_t-\Omega\partial_\phi\right)^2-\partial_\rho^2-\frac{1}{\rho}\partial_\rho-\frac{1}{\rho^2}\partial_\phi^2-\partial_z^2.
\label{PhotonDAlembertianCylindrical}
\eea

\section{Solutions for the wave equation}\label{sec:solutionsofwaveequation}

For the electromagnetic field, however, the corresponding wave equation
in Feynman gauge acts on the vector potential through the covariant
vector Laplacian. The distinction between the scalar and vector cases
arises because the components of $A^\mu$ transform under spatial
rotations. Under rotations around the $z$ axis, the temporal component
and the component along the rotation axis are invariant, whereas the
transverse components transform as a two-dimensional spatial vector.
We therefore separate the vector potential into longitudinal and
transverse components,
\bea
A^\mu=\left(A^0,A^x,A^y,A^z\right),
\label{PhotonVectorPotentialDecomposition}
\eea
with
\bea
A^\mu_{\parallel}=\left(A^0,A^z\right),\qquad
A_\perp=\left(A^x, A^y\right).
\label{PhotonParallelTransverseComponents}
\eea

The transverse components transform nontrivially under rotations around
the $z$ axis. A finite rotation by an angle $\theta$ acts on the
transverse components according to
\bea
A_\perp\longrightarrow R_z(\theta)A_\perp,
\eea
where
\bea
R_z(\theta)=\begin{pmatrix}\cos\theta&-\sin\theta\\\sin\theta&\cos\theta\end{pmatrix}.
\label{PhotonRotationMatrixZ}
\eea

For an infinitesimal rotation, $\theta=\delta\theta$, the rotation matrix
can be expanded as
\bea
R_z(\delta\theta)=\mathbb{1}+\delta\theta\begin{pmatrix}0&-1\\1&0\end{pmatrix}+\mathcal O(\delta\theta^2).
\label{PhotonInfinitesimalRotationMatrixZ}
\eea
Introducing the generator of rotations in the transverse vector
representation through
\bea
R_z(\delta\theta)=\mathbb{1}-\ii\delta\theta\,\hat{\mathcal D}_z+\mathcal O(\delta\theta^2),
\label{PhotonRotationGeneratorDefinition}
\eea
we identify
\bea
\hat{\mathcal D}_z=\begin{pmatrix}0&-\ii\\\ii&0\end{pmatrix},
\label{PhotonTransverseRotationGenerator}
\eea
and the infinitesimal transformation of the transverse vector
components takes the form
\bea
A_\perp\longrightarrow\left(\mathbb{1}-\ii\delta\theta\,\hat{\mathcal D}_z\right)A_\perp.
\label{PhotonTransverseInfinitesimalRotation}
\eea

The eigenvectors of $\hat{\mathcal D}_z$ provide a basis in which the action of
rotations around the $z$ axis is diagonal. This basis is useful for
separating the two transverse circular polarizations, since each
polarization is characterized by a definite eigenvalue of the
transverse rotation generator. We therefore introduce the circular
transverse basis
\bea
\epsilon^\mu_+&=&\frac{1}{\sqrt{2}}\begin{pmatrix}0\\1\\\ii\\0\end{pmatrix},\qquad
\epsilon^\mu_-=\frac{1}{\sqrt{2}}\begin{pmatrix}0\\1\\-\ii\\0\end{pmatrix}.
\label{PhotonCircularBasis}
\eea
These polarization vectors satisfy
\bea
\hat{\mathcal D}_z\boldsymbol{\epsilon}_\pm&=&\pm\boldsymbol{\epsilon}_\pm.
\label{PhotonCircularBasisEigenvalues}
\eea

Then, the transverse vector potential can be decomposed in this
basis as
\bea
A_\perp=A_+\boldsymbol{\epsilon}_++A_-\boldsymbol{\epsilon}_-,
\label{PhotonCircularDecomposition}
\eea
so that the action of the rotation generator on each circular component
is diagonal,
\bea
\hat{\mathcal D}_z A_\perp=A_+\boldsymbol{\epsilon}_+-A_-\boldsymbol{\epsilon}_-.
\label{PhotonDiagonalRotationAction}
\eea

The generator of rotations acting on the complete transverse vector field
contains two distinct contributions. The first one acts on the spatial
dependence of the field and is given by the orbital angular momentum
operator $\hat L_z$. The second one acts on the transverse vector index
and is generated by $\hat{\mathcal D}_z$. Thus, the total generator of
rotations around the $z$ axis is
\bea
\hat J_z=\hat L_z+\hat{\mathcal D}_z.
\label{PhotonTotalAngularMomentum}
\eea
For the circular transverse basis introduced above, the action of
$\hat{\mathcal D}_z$ is diagonal, so when the rotation generator acts on a transverse mode
$A_\pm\boldsymbol{\epsilon}_\pm$, one obtains
\bea
\hat J_z(A_\pm\boldsymbol{\epsilon}_\pm)=\left(\hat L_z\pm1\right)A_\pm\boldsymbol{\epsilon}_\pm.
\label{PhotonTotalGeneratorCircularModes}
\eea

The scalar d'Alembertian introduced in
Eq.~\eqref{PhotonScalarDAlembertianDefinition} is not, in general, the complete differential operator acting on the vector potential. In Feynman gauge, the relevant equation is the covariant vector Laplacian given in Eq.~\eqref{PhotonFeynmanGaugeWaveEquation}. The distinction arises because the covariant derivatives act not only on the spacetime dependence of the field but also on its vector index. 

For a mixed tensor $T^\mu_{\ \beta}$, the covariant derivative is defined as
\bea
\nabla_\alpha T^\mu_{\ \beta}=\partial_\alpha T^\mu_{\ \beta}+\Gamma^\mu_{\alpha\lambda}T^\lambda_{\ \beta}-\Gamma^\lambda_{\alpha\beta}T^\mu_{\ \lambda}.
\label{PhotonCovariantDerivativeMixedTensor}
\eea
Applying this definition to $T^\mu_{\ \beta}=\nabla_\beta A^\mu$, we obtain
\bea
\nabla_\alpha\nabla_\beta A^\mu=\partial_\alpha(\nabla_\beta A^\mu)+\Gamma^\mu_{\alpha\lambda}\nabla_\beta A^\lambda-\Gamma^\lambda_{\alpha\beta}\nabla_\lambda A^\mu,\nn\\
\label{PhotonSecondCovariantDerivativeVector}
\eea
and using Eq.~\eqref{PhotonCovariantDerivativeVector} for the covariant derivative of the vector potential, we have
\bea
&&\nabla_\alpha\nabla_\beta A^\mu\nn\\
&=&\partial_\alpha\partial_\beta A^\mu+\partial_\alpha\left(\Gamma^\mu_{\beta\lambda}A^\lambda\right)+\Gamma^\mu_{\alpha\lambda}\nabla_\beta A^\lambda-\Gamma^\lambda_{\alpha\beta}\nabla_\lambda A^\mu.\nn\\
\label{PhotonSecondCovariantDerivativeExpanded}
\eea

Contracting the derivative indices with the inverse metric, the vector Laplacian becomes
{\small
\bea
&&\nabla_\beta\nabla^\beta A^\mu\nn\\
&=&g^{\alpha\beta}\left[\partial_\alpha\partial_\beta A^\mu+\partial_\alpha\left(\Gamma^\mu_{\beta\lambda}A^\lambda\right)+\Gamma^\mu_{\alpha\lambda}\nabla_\beta A^\lambda-\Gamma^\lambda_{\alpha\beta}\nabla_\lambda A^\mu\right].\nn\\
\label{PhotonVectorLaplacianExpanded}
\eea
}

The connection terms appearing in Eq.~\eqref{PhotonVectorLaplacianExpanded}
are determined by the Christoffel symbols, which for the metric under
consideration are given by Ec~\eqref{PhotonChristoffelSymbols}. Using the rotating metric, the nonvanishing Christoffel symbols are
\bea
\Gamma^x_{tt}&=&-\Omega^2x,\nn\\
\Gamma^y_{tt}&=&-\Omega^2y,\nn\\
\Gamma^x_{ty}&=&\Gamma^x_{yt}=-\Omega,\nn\\
\Gamma^y_{tx}&=&\Gamma^y_{xt}=\Omega.
\label{PhotonNonzeroChristoffelSymbols}
\eea

We first consider the temporal component of the vector potential. From Eq.~\eqref{PhotonNonzeroChristoffelSymbols}, all nonvanishing Christoffel symbols have spatial upper indices. In particular, $\Gamma^0_{\alpha\lambda}=0$. Therefore, the terms in Eq.~\eqref{PhotonVectorLaplacianExpanded} containing an upper temporal Christoffel symbol vanish, and the vector Laplacian for the temporal component reduces to
\bea
\nabla_\beta\nabla^\beta A^0=\Box A^0,
\label{PhotonTemporalVectorLaplacian}
\eea
then, the equation of motion for $A^0$ takes the form
\bea
\Box A^0=0.
\label{PhotonTemporalWaveEquation}
\eea

The same simplification occurs for the component along the rotation axis. Since Eq.~\eqref{PhotonNonzeroChristoffelSymbols} contains no Christoffel symbols with an upper $z$ index, $\Gamma^z_{\alpha\lambda}=0$. Hence, the vector Laplacian acting on $A^z$ becomes
\bea
\nabla_\beta\nabla^\beta A^z=\Box A^z.
\label{PhotonLongitudinalVectorLaplacian}
\eea
and its equation of motion is
\bea
\Box A^z=0.
\label{PhotonLongitudinalWaveEquation}
\eea
Both $A^0$ and $A^z$ are therefore governed by the same differential operator that acts on a scalar field, namely the scalar d'Alembertian given in Eq.~\eqref{PhotonDAlembertianCylindrical}.

The transverse components require a separate treatment because the
rotating coordinate transformation acts nontrivially on the transverse
vector indices. Substituting the nonvanishing Christoffel symbols of
Eq.~\eqref{PhotonNonzeroChristoffelSymbols} into the vector Laplacian of
Eq.~\eqref{PhotonVectorLaplacianExpanded}, the $x$ component becomes
\bea
\nabla_\beta\nabla^\beta A^x&=&\Box A^x+2\Omega^2x\,\partial_yA^y-2\Omega^2y\,\partial_xA^y-\Omega^2A^x\nn\\
&+&2\Omega\,\partial_yA^0-2\Omega\,\partial_tA^y.
\label{PhotonTransverseVectorLaplacianX}
\eea

Similarly, the $y$ component is given by
\bea
\nabla_\beta\nabla^\beta A^y&=&\Box A^y-2\Omega^2x\,\partial_yA^x+2\Omega^2y\,\partial_xA^x-\Omega^2A^y\nn\\
&-&2\Omega\,\partial_xA^0+2\Omega\,\partial_tA^x.
\label{PhotonTransverseVectorLaplacianY}
\eea

The terms beyond the scalar d'Alembertian originate from the connection
terms associated with the transverse vector indices and from the
coupling between the temporal and transverse components induced by the
rotating coordinate basis. To make this structure explicit, we collect
the transverse components into the vector
\bea
A_\perp=\begin{pmatrix}A^x\\A^y\end{pmatrix}.
\label{PhotonTransverseVectorNotation}
\eea

The terms proportional to $\Omega\partial_t$ in
Eqs.~\eqref{PhotonTransverseVectorLaplacianX} and
\eqref{PhotonTransverseVectorLaplacianY} can be expressed in terms of the
transverse rotation generator defined in
Eq.~\eqref{PhotonTransverseRotationGenerator}. Explicitly,
\bea
-2\ii\Omega\hat{\mathcal D}_z\partial_tA_\perp=\begin{pmatrix}-2\Omega\partial_tA^y\\2\Omega\partial_tA^x\end{pmatrix}.
\label{PhotonTransverseTimeRotationTerm}
\eea

The terms containing spatial derivatives of the transverse components
can likewise be expressed in terms of the orbital angular momentum
operator $\hat L_z$ and the transverse rotation generator,
\bea
-2\Omega^2\hat L_z\hat{\mathcal D}_zA_\perp=\begin{pmatrix}2\Omega^2x\,\partial_yA^y-2\Omega^2y\,\partial_xA^y\\-2\Omega^2x\,\partial_yA^x+2\Omega^2y\,\partial_xA^x\end{pmatrix}.
\label{PhotonTransverseOrbitalRotationTerm}
\eea

The remaining term proportional to $\Omega^2$ is obtained from the
square of the transverse rotation generator. Since
\bea
\hat{\mathcal D}_z^2=\mathbb{1},
\label{PhotonTransverseGeneratorSquare}
\eea
one finds
\bea
-\Omega^2\hat{\mathcal D}_z^2A_\perp=-\Omega^2A_\perp.
\label{PhotonTransverseRotationSquareTerm}
\eea

Finally, the terms involving derivatives of the temporal component are
collected as
\bea
2\ii\Omega\hat{\mathcal D}_z\nabla_\perp A^0=\begin{pmatrix}2\Omega\partial_yA^0\\-2\Omega\partial_xA^0\end{pmatrix}.
\label{PhotonTemporalTransverseMixing}
\eea

Combining these contributions with the scalar d'Alembertian, the
transverse vector Laplacian can be written as
\bea
\nabla_\beta\nabla^\beta A_\perp&=&\left[\Box-2\ii\Omega\hat{\mathcal D}_z\partial_t-2\Omega^2\hat L_z\hat{\mathcal D}_z-\Omega^2\hat{\mathcal D}_z^2\right]A_\perp\nn\\
&+&2\ii\Omega\hat{\mathcal D}_z\nabla_\perp A^0.
\label{PhotonTransverseVectorLaplacianCompact}
\eea

The operators $\hat L_z$ and $\hat{\mathcal D}_z$ act on different degrees of
freedom. The former acts on the spatial dependence of the field, whereas
the latter acts on its transverse vector index. Then, they
commute,
\bea
[\hat L_z,\hat{\mathcal D}_z]=0.
\label{PhotonOrbitalSpinGeneratorCommutator}
\eea

Using the rotating d'Alembertian of Eq.~\eqref{PhotonRotatingDAlembertian}
and the relation in Eq.~\eqref{PhotonTransverseGeneratorSquare}, the
terms involving the temporal derivative, the orbital generator, and the
transverse vector generator can be combined as
\bea
&&\Box-2\ii\Omega\hat{\mathcal D}_z\partial_t-2\Omega^2\hat L_z\hat{\mathcal D}_z-\Omega^2\hat{\mathcal D}_z^2\nn\\
&=&\left(\partial_t-\ii\Omega\hat L_z-\ii\Omega\hat{\mathcal D}_z\right)^2-\nabla_\perp^2-\partial_z^2.
\label{PhotonCombinedRotationOperator}
\eea

Introducing the total rotation generator defined in
Eq.~\eqref{PhotonTotalAngularMomentum},
the transverse vector Laplacian therefore takes the form
\bea
&&\nabla_\beta\nabla^\beta A_\perp\nn\\
&=&\left[\left(\partial_t-\ii\Omega\hat J_z\right)^2-\nabla_\perp^2-\partial_z^2\right]A_\perp+2\ii\Omega\hat{\mathcal D}_z\nabla_\perp A^0.\nn\\
\label{PhotonTransverseVectorWaveOperator}
\eea

The rotational operator in Eq.~\eqref{PhotonTransverseVectorWaveOperator}
is governed by the total angular momentum $\hat J_z$, which combines the
orbital contribution associated with the spatial dependence of the field
with the vector contribution associated with the transformation of its
transverse components.

We now construct the cylindrical solutions. We first consider the temporal
and longitudinal sectors, for which the vector Laplacian reduces to the
scalar d'Alembertian. Since the wave operator commutes with the orbital
angular momentum operator $\hat L_z$, the angular dependence of the modes
can be chosen as an eigenfunction of $\hat L_z$. Thus, the temporal and
longitudinal components can be written as
\bea
(A^0)^\mu_\lambda(x)&=&\begin{pmatrix}1\\0\\0\\0\end{pmatrix}R_l(\rho)e^{-\ii Et}e^{\ii l\phi}e^{\ii q_z z},
\label{PhotonParallelMode0}
\eea
and
\bea
(A^z)^\mu_\lambda(x)&=&\begin{pmatrix}0\\0\\0\\1\end{pmatrix}R_l(\rho)e^{-\ii Et}e^{\ii l\phi}e^{\ii q_z z},
\label{PhotonParallelModez}
\eea
where
\bea
\lambda=\{l,E,q_z,|\mathbf{q}_\perp|\},\qquad l\in\mathbb Z.
\label{PhotonModeLabels}
\eea

For these components, $\hat{\mathcal D}_z$ does not contribute to the
rotational generator, so that $\hat J_z=\hat L_z$. Since
\bea
\hat L_z e^{\ii l\phi}=l e^{\ii l\phi},
\label{PhotonOrbitalEigenvalue}
\eea
the action of the rotationally shifted temporal derivative is
\bea
\left(\partial_t-\ii\Omega\hat L_z\right)e^{-\ii Et}e^{\ii l\phi}=-\ii(E+l\Omega)e^{-\ii Et}e^{\ii l\phi}.
\label{PhotonShiftedDerivative}
\eea
It is therefore convenient to define the shifted energy
\bea
\widetilde E_l\equiv E+l\Omega.
\label{PhotonShiftedEnergy}
\eea

Substitution of the above form into the wave equation gives
\bea
\left[-\widetilde E_l^{\,2}-\nabla^2\right]\left[R_l(\rho)e^{\ii l\phi}e^{\ii q_z z}\right]=0.
\label{PhotonModeWaveEquation}
\eea
Using
\bea
\nabla^2=\partial_\rho^2+\frac1\rho\partial_\rho+\frac1{\rho^2}\partial_\phi^2+\partial_z^2,
\label{PhotonCylindricalLaplacian}
\eea
the radial function satisfies
\bea
\left[\frac{d^2}{d\rho^2}+\frac1\rho\frac{d}{d\rho}-\frac{l^2}{\rho^2}+|\mathbf{q}_\perp|^2\right]R_l(\rho)=0,
\label{PhotonRadialBesselEquation}
\eea
where
\bea
|\mathbf{q}_\perp|^2=\widetilde E_l^{\,2}-q_z^2.
\label{PhotonMasslessDispersion}
\eea

This is Bessel's equation, and the solution regular at the origin is
\bea
R_l(\rho)=J_l(|\mathbf{q}_\perp|\rho).
\label{PhotonRegularRadialSolution}
\eea

For the transverse sector, it is convenient to label the modes by their
orbital angular momentum eigenvalue $l$. According to
Eq.~\eqref{PhotonTotalGeneratorCircularModes}, the eigenvalues of the total
angular momentum operator for circularly polarized modes follow from the
orbital eigenvalue $l$ and the vector contribution $\pm1$, yielding
$j_\pm=l\pm1$. The angular dependence of the corresponding modes is therefore fixed by
\bea
\hat L_z e^{\ii l\phi}=l e^{\ii l\phi}.
\label{PhotonCircularOrbitalEigenvalue}
\eea
We can Then write the transverse circular modes as
\bea
(A^\pm)^\mu_\lambda(x)=R^\pm_l(\rho)e^{-\ii Et}e^{\ii l\phi}e^{\ii q_z z}\boldsymbol{\epsilon}_\pm^\mu.
\label{PhotonCircularOrbitalModes}
\eea

The action of the total rotational generator on these modes is
\bea
\hat J_z(A^\pm)^\mu_\lambda=(l\pm1)(A^\pm)^\mu_\lambda.
\label{PhotonCircularTotalAngularMomentumModeEigenvalue}
\eea
Then, the rotationally shifted temporal derivative appearing in
Eq.~\eqref{PhotonTransverseVectorWaveOperator} acts on these modes as
\bea
(\partial_t-\ii\Omega\hat J_z)(A^\pm)^\mu_\lambda=-\ii[E+\Omega(l\pm1)](A^\pm)^\mu_\lambda.
\label{PhotonCircularTotalShiftedDerivative}
\eea

For each circular polarization, it is therefore convenient to introduce the corresponding shifted energy
\bea
\widetilde E_{l,\pm}\equiv E+\Omega(l\pm1).
\label{PhotonCircularTotalShiftedEnergy}
\eea

The transverse wave equation contains a homogeneous part acting on $A_\perp$ and an inhomogeneous contribution generated by the temporal component $A^0$. We therefore solve these two contributions separately and construct the complete transverse solution by superposition. In particular, we first consider the homogeneous equation obtained by setting $A^0=0$ in Eq.~\eqref{PhotonTransverseVectorWaveOperator},
\bea
\left[\left(\partial_t-\ii\Omega\hat J_z\right)^2-\nabla_\perp^2-\partial_z^2\right]A_\perp^{\text{hom.}}=0.
\label{PhotonTransverseHomogeneousWaveEquation}
\eea

Since the homogeneous operator is diagonal in the eigenbasis of $\hat J_z$, the two circular polarization sectors can be solved independently. Using Eq.~\eqref{PhotonCircularOrbitalModes} and the corresponding total angular momentum eigenvalues,
\bea
\hat J_z(A_l^\pm)^\mu=(l\pm1)(A_l^\pm)^\mu,
\eea
the radial equations for the two circular modes can be written as a single equation,
\bea
\left[\frac{d^2}{d\rho^2}+\frac{1}{\rho}\frac{d}{d\rho}-\frac{l^2}{\rho^2}+|\mathbf q_\perp|^2\right]R_l^\pm(\rho)=0,
\label{PhotonHomogeneousRadialEquation}
\eea
where the transverse momentum magnitude is chosen to be the same for both modes. The corresponding regular radial solutions are
\bea
R_l^\pm(\rho)=C_l^\pm J_l(|\mathbf q_\perp|\rho).
\eea
The frequency entering the radial equation is the shifted energy
\bea
\widetilde E_{l,\pm}=E_{l,\pm}+\Omega(l\pm1),
\eea
which gives
\bea
\widetilde E_{l,+}=\sqrt{|\mathbf q_\perp|^2+q_z^2}=\widetilde E_{l,-},
\eea
while the corresponding frequencies in the rotating coordinates are
\bea
E_{l,+}&=&\sqrt{|\mathbf q_\perp|^2+q_z^2}-\Omega(l+1),\nn\\ E_{l,-}&=&\sqrt{|\mathbf q_\perp|^2+q_z^2}-\Omega(l-1).
\eea
Thus, the two circular modes have the same transverse momentum magnitude and the same shifted energy, whereas their rotating-frame frequencies differ according to their different total angular momentum eigenvalues.

For real $|\mathbf q_\perp|$, the solutions regular at the origin are
\bea
R_l^\pm(\rho)=C_l^\pm J_l(|\mathbf q_\perp|\rho),
\label{PhotonHomogeneousRadialSolutions}
\eea
and the general regular homogeneous transverse solution is
\bea
A_{\perp,l}^{\text{hom.}}&=&\Big[C_l^+J_l(|\mathbf q_\perp|\rho)e^{\ii l\phi}\boldsymbol{\epsilon}_++C_l^-J_l(|\mathbf q_\perp|\rho)e^{\ii l\phi}\boldsymbol{\epsilon}_-\Big]\nn\\
&\times&e^{-\ii Et}e^{\ii q_z z}.
\label{PhotonGeneralHomogeneousTransverseSolution}
\eea

We now consider the inhomogeneous contribution generated by the temporal component. From Eq.~\eqref{PhotonRegularRadialSolution}, a temporal mode with orbital angular momentum $l$ has the form of Eq.~\eqref{PhotonParallelMode0}. Since the temporal component has no transverse vector contribution, its total angular momentum is equal to its orbital angular momentum, so that
\bea
\hat J_z(A^0)^\mu_\lambda=l(A^0)^\mu_\lambda.
\label{PhotonTemporalModeTotalAngularMomentum}
\eea
The temporal component therefore belongs to the total angular momentum sector $j=l$. The temporal component enters the transverse equation through the mixing term already identified in Eq.~\eqref{PhotonTemporalTransverseMixing}. Therefore, the inhomogeneous transverse equation is
\bea
\left[\left(\partial_t-\ii\Omega\hat J_z\right)^2-\nabla_\perp^2-\partial_z^2\right]A_\perp^{\text{part.}}=-2\ii\Omega\hat{\mathcal D}_z\nabla_\perp A^0.\nn\\
\label{PhotonTransverseInhomogeneousWaveEquation}
\eea

The transverse gradient of the temporal mode can be evaluated in the circular basis. Using the cylindrical representation of the transverse gradient and the Bessel recurrence relations, one obtains
{\small
\bea
&&\nabla_\perp\left[J_l(|\mathbf q_\perp|\rho)e^{\ii l\phi}\right]\nn\\
&=&\frac{|\mathbf q_\perp|}{\sqrt{2}}\left[J_{l-1}(|\mathbf q_\perp|\rho)e^{\ii(l-1)\phi}\boldsymbol{\epsilon}_+-J_{l+1}(|\mathbf q_\perp|\rho)e^{\ii(l+1)\phi}\boldsymbol{\epsilon}_-\right].\nn\\
\label{PhotonTemporalModeTransverseGradientBessel}
\eea
}

The two terms in Eq.~\eqref{PhotonTemporalModeTransverseGradientBessel}
have different orbital angular momenta because the transverse gradient
changes the orbital content of the original mode with angular momentum
$l$. Using Eq.~\eqref{PhotonCircularBasisEigenvalues}, the first term has
orbital angular momentum $l-1$ and vector contribution $+1$, yielding
\bea
j=(l-1)+1=l,
\eea
whereas the second term has orbital angular momentum $l+1$ and vector
contribution $-1$, yielding
\bea
j=(l+1)-1=l.
\eea
Thus, the transverse gradient decomposes the mode into two circular
components with different orbital angular momenta, while preserving the
total angular momentum $j=l$. Applying $\hat{\mathcal D}_z$ gives
\bea
2\ii\Omega\hat{\mathcal D}_z\nabla_\perp A^0&=&\sqrt{2}\ii\Omega|\mathbf q_\perp|a_l\Big[J_{l-1}(|\mathbf q_\perp|\rho)e^{\ii(l-1)\phi}\boldsymbol{\epsilon}_+\nn\\
&+&J_{l+1}(|\mathbf q_\perp|\rho)e^{\ii(l+1)\phi}\boldsymbol{\epsilon}_-\Big]e^{-\ii Et}e^{\ii q_z z}.\nn\\
\label{PhotonTemporalModeTransverseSource}
\eea

The two terms in Eq.~\eqref{PhotonTemporalModeTransverseSource} have
different orbital angular momenta as a consequence of the action of
$\nabla_\perp$ on the original mode with orbital angular momentum $l$.
For the $\boldsymbol{\epsilon}_+$ component, the orbital contribution
is $l-1$ and the vector contribution is $+1$, whereas for the
$\boldsymbol{\epsilon}_-$ component they are $l+1$ and $-1$,
respectively. Hence, both components have the same total angular
momentum,
\bea
j_+=(l-1)+1=l,\qquad j_-=(l+1)-1=l.\nn\\
\eea
Thus, the transverse source generated by the temporal mode remains
entirely within the total angular momentum sector $j=l$. In this sense,
the label $l$ is inherited directly from the original orbital
angular-momentum eigenstate, while the action of $\nabla_\perp$
redistributes its orbital angular momentum between the two circular
components without changing the total angular momentum.

The structure of the inhomogeneous wave operator further justifies
constructing the particular solution within this same sector. Since the
operator is diagonal in the eigenvalues associated with the temporal,
longitudinal, and total angular momentum dependences, its action on a
mode proportional to
$e^{-\ii Et}e^{\ii j\phi}e^{\ii q_z z}$ preserves the corresponding
values of $E$, $j$, and $q_z$. Its transverse part acts nontrivially only
on the radial dependence and on the orbital angular momentum of each
circular component. Therefore, the operator does not generate new
temporal, longitudinal, or total angular momentum dependences, and the
particular solution can be constructed with the same dependence as the
source, while leaving the radial functions undetermined.

We therefore take the particular solution in the form
\bea
&&(A_{\perp,l}^{\text{part.}})^\mu\nn\\
&=&\left[\tilde{R}_{l}^+(\rho)e^{\ii(l-1)\phi}\boldsymbol{\epsilon}_+^\mu+\tilde{R}_{l}^-(\rho)e^{\ii(l+1)\phi}\boldsymbol{\epsilon}_-^\mu\right]e^{-\ii Et}e^{\ii q_z z}.\nn\\
\label{PhotonTransverseParticularAnsatz}
\eea
Here, $\tilde{R}_{l}^\pm(\rho)$ are unknown radial functions that are
determined by substituting the ansatz into the inhomogeneous transverse
wave equation.

Since the total angular momentum of the source is $j=l$, the shifted energy entering the radial equations is
\bea
\widetilde E_l=E+l\Omega.
\label{PhotonParticularShiftedEnergy}
\eea
Substitution of Eq.~\eqref{PhotonTransverseParticularAnsatz} into Eq.~\eqref{PhotonTransverseInhomogeneousWaveEquation} gives
\bea
&&\left[\frac{d^2}{d\rho^2}+\frac{1}{\rho}\frac{d}{d\rho}-\frac{(l-1)^2}{\rho^2}+|\mathbf q_\perp|^2\right]\tilde{R}_{l}^+(\rho)\nn\\
&=&-\sqrt{2}\ii\Omega|\mathbf q_\perp|a_lJ_{l-1}(|\mathbf q_\perp|\rho),
\label{PhotonParticularRadialEquationPlus}
\eea

and
\bea
&&\left[\frac{d^2}{d\rho^2}+\frac{1}{\rho}\frac{d}{d\rho}-\frac{(l+1)^2}{\rho^2}+|\mathbf q_\perp|^2\right]\tilde{R}_{l}^-(\rho)\nn\\
&=&-\sqrt{2}\ii\Omega|\mathbf q_\perp|a_lJ_{l+1}(|\mathbf q_\perp|\rho).
\label{PhotonParticularRadialEquationMinus}
\eea
It is useful to define the radial operator
\bea
\mathcal L_n=\frac{d^2}{d\rho^2}+\frac{1}{\rho}\frac{d}{d\rho}-\frac{n^2}{\rho^2}+|\mathbf q_\perp|^2.
\label{PhotonParticularBesselOperator}
\eea

The required particular solutions can be obtained by considering $\rho J_l(|\mathbf q_\perp|\rho)$. Using the Bessel equation satisfied by $J_l(|\mathbf q_\perp|\rho)$, one finds
\bea
\mathcal L_{l-1}\left[\rho J_l(|\mathbf q_\perp|\rho)\right]&=&2|\mathbf q_\perp|J_{l-1}(|\mathbf q_\perp|\rho),\nn\\
\mathcal L_{l+1}\left[\rho J_l(|\mathbf q_\perp|\rho)\right]&=&-2|\mathbf q_\perp|J_{l+1}(|\mathbf q_\perp|\rho).
\label{PhotonParticularBesselIdentities}
\eea
The first relation in Eq.~\eqref{PhotonParticularBesselIdentities} gives the particular solution for the positive circular component,
\bea
\tilde{R}_{l}^+(\rho)=-\frac{\ii\Omega a_l}{\sqrt{2}}\rho J_l(|\mathbf q_\perp|\rho),
\label{PhotonParticularSolutionPlus}
\eea
while the second relation gives
\bea
\tilde{R}_{l}^-(\rho)=\frac{\ii\Omega a_l}{\sqrt{2}}\rho J_l(|\mathbf q_\perp|\rho).
\label{PhotonParticularSolutionMinus}
\eea

The particular transverse field is therefore
\bea
A_{\perp,l}^{\text{part.}}&=&\frac{\ii\Omega a_l}{\sqrt{2}}\rho J_l(|\mathbf q_\perp|\rho)e^{-\ii Et}e^{\ii q_z z}\nn\\
&\times&\left[-e^{\ii(l-1)\phi}\boldsymbol{\epsilon}_++e^{\ii(l+1)\phi}\boldsymbol{\epsilon}_-\right].
\label{PhotonTransverseParticularSolution}
\eea

The complete transverse solution is obtained by adding this particular solution to the general homogeneous solution
\bea
A_{\perp,l}^\mu=(A_{\perp,l}^{\mathrm{hom.}})^\mu+(A_{\perp,l}^{\mathrm{part.}})^\mu.
\label{PhotonCompleteTransverseSolution}
\eea

\section{The Physical Photon Modes}\label{subsec:PhotonModes}

\subsection{Construction of the physical photon modes}\label{secVA}

It is convenient to organize the photon modes according to the eigenvalue of the total angular momentum operator of Eq.~\eqref{PhotonTotalAngularMomentum} and therefore we use $j$ as a common label for all modes belonging to the same total angular momentum sector,
\bea
&&\hat J_z A_j^\mu=jA_j^\mu.
\eea

With this convention, $j$ does not denote the orbital angular momentum of each component. Instead, the orbital angular momentum is $j$ for the temporal and longitudinal components, $j-1$ for the $\boldsymbol{\epsilon}_+$ component, and $j+1$ for the $\boldsymbol{\epsilon}_-$ component. Thus,
\bea
A_j^0&\sim& J_j(|\mathbf q_\perp|\rho)e^{\ii j\phi},\nn\\ A_j^z&\sim& J_j(|\mathbf q_\perp|\rho)e^{\ii j\phi},\nn\\
A_j^+&\sim& J_{j-1}(|\mathbf q_\perp|\rho)e^{\ii(j-1)\phi}\boldsymbol{\epsilon}_+,\nn\\
A_j^-&\sim& J_{j+1}(|\mathbf q_\perp|\rho)e^{\ii(j+1)\phi}\boldsymbol{\epsilon}_-.
\eea
Indeed,
\bea
\hat J_z\left[e^{\ii(j-1)\phi}\boldsymbol{\epsilon}_+\right]&=&je^{\ii(j-1)\phi}\boldsymbol{\epsilon}_+,\nn\\
\hat J_z\left[e^{\ii(j+1)\phi}\boldsymbol{\epsilon}_-\right]&=&je^{\ii(j+1)\phi}\boldsymbol{\epsilon}_-.
\eea
Hence, all four components belong to the same $j$ sector despite having different orbital angular momenta.

Having fixed the gauge, the equation of motion is
\bea
&&\nabla_\mu\nabla^\mu A^\nu=0.
\eea
We now determine the physical photon modes by analyzing the solutions of this equation within each fixed-$j$ sector. The four coefficients $a_j$, $C_j^+$, $C_j^-$, and $C_j^z$ which accompany each solution $(A^0,A_\perp,A^z)$. The physical content is then obtained by imposing the Lorenz condition and identifying the remaining redundancy associated with gauge transformations that preserve the gauge-fixed equation of motion. 

In the $j$ sector, such residual transformations are generated by
\bea
&&A^\mu\rightarrow A^\mu+\nabla^\mu\Lambda.
\label{residual}
\eea
Since the Feynman gauge condition gives $\nabla_\mu A^\mu=0$, the transformed field must satisfy
\bea
&&\nabla_\mu A'^\mu=\nabla_\mu A^\mu+\nabla_\mu\nabla^\mu\Lambda=0,
\eea
which, using $\nabla_\mu A^\mu=0$, requires the gauge function to satisfy
\bea
&&\Box\Lambda\equiv\nabla_\mu\nabla^\mu\Lambda=0.
\eea
Thus, the residual gauge transformations that preserve the Feynman gauge are
\bea
&&A^\mu\rightarrow A^\mu+\nabla^\mu\Lambda,\qquad \Box\Lambda=0.
\eea

We choose the residual gauge parameter as a regular scalar mode of the same fixed-$j$ sector. Since $\Lambda$ is a scalar, its total angular momentum is entirely orbital, and therefore an eigenmode with $\hat J_z\Lambda=j\Lambda$ has the angular dependence $e^{\ii j\phi}$. Requiring the same energy $E$ and longitudinal momentum $q_z$ as the photon modes, and choosing the regular radial solution of $\Box\Lambda=0$, we take
\bea
&&\Lambda=\lambda_jJ_j(|\mathbf q_\perp|\rho)e^{-\ii Et}e^{\ii j\phi}e^{\ii q_z z}.
\eea
The divergence of the temporal component is
\bea
&&\nabla_\mu A^{0\mu}=-\ii\widetilde E_j a_jJ_j(|\mathbf q_\perp|\rho)e^{-\ii Et}e^{\ii j\phi}e^{\ii q_z z},\nn\\
\eea
while the longitudinal component gives
\bea
&&\nabla_\mu A^{z\mu}=\ii q_zC_j^zJ_j(|\mathbf q_\perp|\rho)e^{-\ii Et}e^{\ii j\phi}e^{\ii q_z z}.\nn\\
\eea

The divergence of the homogeneous transverse contribution is
\bea
&&\nabla_\mu(A_{\perp,j}^{\rm hom})^\mu=\frac{|\mathbf q_\perp|}{\sqrt{2}}(C_j^--C_j^+)J_j(|\mathbf q_\perp|\rho)e^{-\ii Et}e^{\ii j\phi}e^{\ii q_z z},\nn\\
\eea
whereas the particular solution contributes
\bea
&&\nabla_\mu(A_{\perp,j}^{\rm part})^\mu=-\ii j\Omega a_jJ_j(|\mathbf q_\perp|\rho)e^{-\ii Et}e^{\ii j\phi}e^{\ii q_z z}.\nn\\
\eea
Therefore,
\bea
\nabla_\mu A^\mu&=&\left[\frac{|\mathbf q_\perp|}{\sqrt{2}}(C_j^--C_j^+)+\ii q_zC_j^z-\ii\widetilde E_ja_j\right]\nn\\
&\times&J_j(|\mathbf q_\perp|\rho)e^{-\ii Et}e^{\ii j\phi}e^{\ii q_z z}.
\eea

The Lorenz condition Then imposes
\bea
&&\frac{|\mathbf q_\perp|}{\sqrt{2}}(C_j^--C_j^+)+\ii q_zC_j^z-\ii\widetilde E_ja_j=0.
\label{PhotonLorenzConstraint}
\eea

The residual gauge transformation modifies the coefficients of the four components of the field via Eq.~\eqref{residual}, so that for the temporal component, using the definition of $\widetilde E_j$ introduced above, we have
\bea
&&\nabla^0\Lambda=-\ii\widetilde E_j\lambda_jJ_j(|\mathbf q_\perp|\rho)e^{-\ii Et}e^{\ii j\phi}e^{\ii q_z z}.
\eea
Comparing this result with the temporal mode proportional to $a_j$, the transformed coefficient is
\bea
&&a'_j=a_j-\ii\widetilde E_j\lambda_j.
\eea
For the transverse components, the gradient of the scalar mode is
\bea
\nabla_\perp\Lambda&=&\frac{|\mathbf q_\perp|\lambda_j}{\sqrt{2}}\Big[J_{j-1}(|\mathbf q_\perp|\rho)e^{\ii(j-1)\phi}\boldsymbol{\epsilon}_+\nn\\
&-&J_{j+1}(|\mathbf q_\perp|\rho)e^{\ii(j+1)\phi}\boldsymbol{\epsilon}_-\Big]e^{-\ii Et}e^{\ii q_z z}.
\eea

The first term has the same mode structure as the $C_j^+$ contribution, while the second has the same mode structure as the $C_j^-$ contribution. Therefore,
\bea
&&C_j^{+\prime}=C_j^+-\frac{|\mathbf q_\perp|}{\sqrt{2}}\lambda_j,
\eea
and
\bea
&&C_j^{-\prime}=C_j^-+\frac{|\mathbf q_\perp|}{\sqrt{2}}\lambda_j.
\eea

For the longitudinal component, since $g^{zz}=-1$, we obtain
\bea
&&\nabla^z\Lambda=-\partial_z\Lambda=-\ii q_z\lambda_jJ_j(|\mathbf q_\perp|\rho)e^{-\ii Et}e^{\ii j\phi}e^{\ii q_z z}.\nn\\
\eea
Comparing with the longitudinal mode proportional to $C_j^z$ gives
\bea
&&C_j^{z\prime}=C_j^z-\ii q_z\lambda_j.
\eea

Thus, the residual gauge transformation changes the four coefficients according to
\bea
&&a'_j=a_j-\ii\widetilde E_j\lambda_j,\qquad C_j^{+\prime}=C_j^+-\frac{|\mathbf q_\perp|}{\sqrt{2}}\lambda_j,\nn\\
o&&C_j^{-\prime}=C_j^-+\frac{|\mathbf q_\perp|}{\sqrt{2}}\lambda_j,\qquad C_j^{z\prime}=C_j^z-\ii q_z\lambda_j.\nn\\
\eea

For $\widetilde E_j\neq0$, choosing
\bea
&&\lambda_j=\frac{\ii a_j}{\widetilde E_j}
\eea
sets $a_j=0$. 

The Lorenz condition then becomes
\bea
&&\frac{|\mathbf q_\perp|}{\sqrt{2}}(C_j^--C_j^+)+\ii q_zC_j^z=0.
\label{PhotonPhysicalConstraint}
\eea
which leaves a three-dimensional coefficient space subject to one linear constraint. After fixing the residual gauge freedom by setting $a_j=0$, this constraint reduces the remaining solution space to two independent dimensions. We can therefore choose any two linearly independent solutions of Eq.~\eqref{PhotonPhysicalConstraint} as a basis for the physical modes. A convenient choice is
\bea
&&(C_j^+,C_j^-,C_j^z)=(1,1,0),
\label{def:v1}
\eea
which corresponds to a purely transverse mode, and
\bea
&&(C_j^+,C_j^-,C_j^z)=(q_z,-q_z,-\ii\sqrt{2}|\mathbf q_\perp|),
\label{def:v2}
\eea
which contains both transverse and longitudinal components.

The first vector satisfies the Lorenz condition because $C_j^-=C_j^+$ and $C_j^z=0$. Using it as the coefficients of the general solution gives
\bea
&&(A_{j,1})^\mu=\left(A_j^+\right)^\mu+\left(A_j^-\right)^\mu.
\eea
Substituting the corresponding transverse solutions, we obtain
\bea
(A_{j,1})^\mu&=&\mathcal N_1\Big[J_{j-1}(|\mathbf q_\perp|\rho)e^{\ii(j-1)\phi}\boldsymbol{\epsilon}_+^\mu\nn\\
&+&J_{j+1}(|\mathbf q_\perp|\rho)e^{\ii(j+1)\phi}\boldsymbol{\epsilon}_-^\mu\Big]e^{-\ii Et}e^{\ii q_z z}.\nn\\
\eea

For the second vector, the coefficients are
\bea
&&C_j^+=q_z,\qquad C_j^-=-q_z,\qquad C_j^z=-\ii\sqrt{2}|\mathbf q_\perp|.\nn\\
\eea

Using these coefficients in the general solution gives
\bea
(A_{j,2})^\mu=q_z\left(A_j^+\right)^\mu-q_z\left(A_j^-\right)^\mu-\ii\sqrt{2}|\mathbf q_\perp|\left(A_j^z\right)^\mu.\nn\\
\eea

Substituting the corresponding transverse and longitudinal solutions then gives
   \bea
(A_{j,2})^\mu&=&\mathcal N_2\Big[q_zJ_{j-1}(|\mathbf q_\perp|\rho)e^{\ii(j-1)\phi}\boldsymbol{\epsilon}_+^\mu\nn\\
&-&q_zJ_{j+1}(|\mathbf q_\perp|\rho)e^{\ii(j+1)\phi}\boldsymbol{\epsilon}_-^\mu\nn\\
&-&\ii\sqrt{2}|\mathbf q_\perp|J_j(|\mathbf q_\perp|\rho)e^{\ii j\phi}\delta^\mu_z\Big]e^{-\ii Et}e^{\ii q_z z}.\nn\\
\eea 

And provided that for each sector $j=l$, we get
\begin{widetext}
    \bea
A_{l,1}^\mu&=&\mathcal N_1\Big[J_{l-1}(|\mathbf q_\perp|\rho)e^{\ii(l-1)\phi}\boldsymbol{\epsilon}_+^\mu+J_{l+1}(|\mathbf q_\perp|\rho)e^{\ii(l+1)\phi}\boldsymbol{\epsilon}_-^\mu\Big]e^{-\ii Et}e^{\ii q_z z}.
\label{PhotonModeOneFinal}
\eea
  \bea
A_{l,2}^\mu&=&\mathcal N_2\Big[q_zJ_{l-1}(|\mathbf q_\perp|\rho)e^{\ii(l-1)\phi}\boldsymbol{\epsilon}_+^\mu-q_zJ_{l+1}(|\mathbf q_\perp|\rho)e^{\ii(l+1)\phi}\boldsymbol{\epsilon}_-^\mu -\ii\sqrt{2}|\mathbf q_\perp|J_l(|\mathbf q_\perp|\rho)e^{\ii l\phi}\delta^\mu_z\Big]e^{-\ii Et}e^{\ii q_z z}.
\label{PhotonModeTwoFinal}
\eea 
\end{widetext}

The residual gauge transformation can be used to set $a_l=0$ provided that $\widetilde E_l\neq0$. Indeed, the temporal coefficient transforms according to $a_l'=a_l-\ii\widetilde E_l\lambda_l$, so that choosing $\lambda_l=\ii a_l/\widetilde E_l$ removes the temporal coefficient. The case $\widetilde E_l=0$ is singular in this construction and must be treated separately. From the dispersion relation for the photon modes, $\widetilde E_l^2=|\mathbf q_\perp|^2+q_z^2$, this condition requires $|\mathbf q_\perp|=q_z=0$. Hence, $\widetilde E_l=0$ corresponds to the zero momentum limit rather than to a propagating photon mode. In this limit, the Lorenz constraint becomes identically satisfied and the residual transformation generated by the scalar mode does not change any of the coefficients. The polarization decomposition therefore becomes degenerate, since there is no nonzero spatial momentum with respect to which a transverse polarization can be defined. We therefore restrict the physical mode construction to propagating modes with $\mathbf q\neq0$, for which $\widetilde E_l>0$ and the residual gauge freedom can be used to impose $a_l=0$.

With the physical modes thus restricted to propagating solutions, we next specify the radial domain and the corresponding spectral structure. The radial equation is a Bessel equation, whose general solutions are linear combinations of $J_n(|\mathbf q_\perp|\rho)$ and $Y_n(|\mathbf q_\perp|\rho)$. Regularity at the origin excludes the Bessel functions of the second kind, leaving the regular solutions $J_n(|\mathbf q_\perp|\rho)$. We consider these solutions on the unbounded transverse plane, $0\leq\rho<\infty$, without introducing a rigid cylindrical boundary at a finite radius $R$. In particular, no condition such as $J_n(|\mathbf q_\perp| R)=0$ is imposed, and the transverse momentum $q_\perp$ remains continuous. This is reflected in the orthogonality relation of the regular Bessel functions, which yields a Dirac delta function in $|\mathbf q_\perp|$ rather than a discrete Kronecker delta. The corresponding continuous measure in $q_\perp$ will be used in the quantization of the field and in the construction of the propagator.

At large radial distances, the regular solutions have the asymptotic form for $|\mathbf q_\perp|\rho\gg1$
\bea
J_n(|\mathbf q_\perp|\rho)\sim\sqrt{\frac{2}{\pi |\mathbf q_\perp|\rho}}\cos\left(|\mathbf q_\perp|\rho-\frac{n\pi}{2}-\frac{\pi}{4}\right),
\eea
therefore, the radial dependence becomes oscillatory at large $\rho$, with an amplitude that decreases as $\rho^{-1/2}$. The modes are therefore not subject to an additional radial boundary condition at large distances, but retain the cylindrical-wave behavior associated with the unbounded transverse domain. As discussed in Sec.~\ref{sec:geometry_and_verbein}, in practice we consider that the rigidly rotating system is confined within a radius $R$, with $R$ satisfying $\Omega R<1$.

The two physical modes in Eqs.~\eqref{PhotonModeOneFinal} and \eqref{PhotonModeTwoFinal} provide a convenient basis for the physical solution space. However, their explicit forms are not symmetric with respect to the three vector structures entering the spatial part of the photon field. In particular, the mode $A_{l,1}^\mu$ contains only the two circular transverse structures, whereas $A_{l,2}^\mu$ contains, in addition, the component proportional to $\delta^\mu_z$. This apparent asymmetry is a property of the particular basis chosen for the physical solution space and does not represent a physical distinction between the two independent photon modes.

In order to symmetrize the vector structure of the photon states, it is convenient to introduce the magnitude of the spatial photon momentum
\bea
|\mathbf q|=\sqrt{|\mathbf q_\perp|^2+q_z^2},
\eea
so that we define
\bea
\boldsymbol e_1=\frac{1}{\sqrt{2}}\boldsymbol v_1,\qquad \boldsymbol e_2=\frac{1}{\sqrt{2}|\mathbf q|}\boldsymbol v_2,
\eea
where $\boldsymbol v_1$ and $\boldsymbol v_2$ are given by Eqs.~\eqref{def:v1} and~\eqref{def:v2} respectively. 

We can then define a new basis through the symmetric and antisymmetric combinations%
\bea
\widetilde{\boldsymbol e}_1=\frac{\boldsymbol e_1+\boldsymbol e_2}{\sqrt{2}},\qquad \widetilde{\boldsymbol e}_2=\frac{\boldsymbol e_1-\boldsymbol e_2}{\sqrt{2}},
\label{PhotonSymmetricPolarizationBasis}
\eea
so that the resulting coefficient vectors are
\bea
\widetilde{\boldsymbol e}_1=\frac{1}{2}\left(1+\frac{q_z}{|\mathbf q|},1-\frac{q_z}{|\mathbf q|},-\ii\sqrt{2}\frac{|\mathbf q_\perp|}{|\mathbf q|}\right),
\eea
\bea
\widetilde{\boldsymbol e}_2=\frac{1}{2}\left(1-\frac{q_z}{|\mathbf q|},1+\frac{q_z}{|\mathbf q|},+\ii\sqrt{2}\frac{|\mathbf q_\perp|}{|\mathbf q|}\right).
\eea

Identifying the three entries with the coefficients multiplying $\boldsymbol{\epsilon}_+^\mu$, $\boldsymbol{\epsilon}_-^\mu$ and $\delta^\mu_z$, respectively, we obtain
\bea
\widetilde C_1^+&=&\frac{1}{2}\left(1+\frac{q_z}{|\mathbf q|}\right),\nn\\
\widetilde C_1^-&=&\frac{1}{2}\left(1-\frac{q_z}{|\mathbf q|}\right),\nn\\
\widetilde C_1^z&=&-\frac{\ii}{\sqrt{2}}\frac{|\mathbf q_\perp|}{|\mathbf q|},
\eea
and
\bea
\widetilde C_2^+&=&\frac{1}{2}\left(1-\frac{q_z}{|\mathbf q|}\right),\nn\\
\widetilde C_2^-&=&\frac{1}{2}\left(1+\frac{q_z}{|\mathbf q|}\right),\nn\\ 
\widetilde C_2^z&=&+\frac{\ii}{\sqrt{2}}\frac{|\mathbf q_\perp|}{|\mathbf q|}.
\eea
The symmetry between the two modes is now explicit, since%
\bea
\widetilde C_2^+=\widetilde C_1^-,\qquad \widetilde C_2^-=\widetilde C_1^+,\qquad \widetilde C_2^z=-\widetilde C_1^z,
\eea
and the two physical modes can therefore be written in the common form, for $\beta=1,2$, as
\begin{widetext}
 \bea
\widetilde A_{l,\beta}^\mu=\widetilde{\mathcal N}_\beta\Big[\widetilde C_\beta^+J_{l-1}(|\mathbf q_\perp|\rho)e^{\ii(l-1)\phi}\boldsymbol{\epsilon}_+^\mu+\widetilde C_\beta^-J_{l+1}(|\mathbf q_\perp|\rho)e^{\ii(l+1)\phi}\boldsymbol{\epsilon}_-^\mu+\widetilde C_\beta^zJ_l(|\mathbf q_\perp|\rho)e^{\ii l\phi}\delta^\mu_z\Big]e^{-\ii Et}e^{\ii q_z z}.
\label{PhotonSymmetricPhysicalModes}
\eea   
\end{widetext}

In this representation, both physical modes contain the same three spatial and vector structures, while their distinction is entirely encoded in the coefficients $\widetilde C_\beta^+$, $\widetilde C_\beta^-$ and $\widetilde C_\beta^z$. The appearance of $\delta^\mu_z$ in both modes does not introduce an additional physical polarization. The physical solution space remains two dimensional, and the new modes are obtained only by a change of basis within this space.

The coefficients defining the new basis satisfy the same reduced Lorenz condition. In particular, for the first mode,
\bea
\frac{|\mathbf q_\perp|}{\sqrt{2}}\left(\widetilde C_1^--\widetilde C_1^+\right)+\ii q_z\widetilde C_1^z=0,
\eea
while for the second mode,
\bea
\frac{|\mathbf q_\perp|}{\sqrt{2}}\left(\widetilde C_2^--\widetilde C_2^+\right)+\ii q_z\widetilde C_2^z=0.
\eea
Thus, the new basis remains entirely within the same two-dimensional physical solution space defined by the reduced Lorenz condition.

{

\subsection{Alternative derivation of the physical photon modes}\label{secVB}

The solutions of the Maxwell equation can be constructed by exploiting the eigenfunctions of the vector Laplacian. Let
\bea
\Psi_a^\mu=e_a^\mu u_j,
\eea
where $e_a^\mu$ denotes a set of vector basis elements satisfying $\nabla_\nu e_a^\mu=0$, and $u_j$ is a scalar eigenfunction with definite angular dependence. To determine the action of the vector Laplacian on $\Psi_a^\mu$, we first consider its covariant derivative. Using
\bea
\nabla_\beta\Psi_a^\mu=\partial_\beta\Psi_a^\mu+\Gamma^\mu_{\beta\lambda}\Psi_a^\lambda,
\eea
we obtain
\bea
\nabla_\beta\Psi_a^\mu&=&\partial_\beta(e_a^\mu u_j)+\Gamma^\mu_{\beta\lambda}e_a^\lambda u_j\nn\\
&=&e_a^\mu\partial_\beta u_j+u_j\left(\partial_\beta e_a^\mu+\Gamma^\mu_{\beta\lambda}e_a^\lambda\right).
\eea
The term in parentheses is the covariant derivative of the vector basis,
\bea
\partial_\beta e_a^\mu+\Gamma^\mu_{\beta\lambda}e_a^\lambda=\nabla_\beta e_a^\mu,
\eea
which vanishes by construction. Therefore,
\bea
\nabla_\beta\Psi_a^\mu=e_a^\mu\partial_\beta u_j.
\eea
Applying a second covariant derivative, we have
\bea
\nabla_\gamma\nabla_\beta\Psi_a^\mu=\nabla_\gamma(e_a^\mu\partial_\beta u_j)=e_a^\mu\nabla_\gamma\partial_\beta u_j+(\nabla_\gamma e_a^\mu)\partial_\beta u_j.\nn\\
\eea
Again, the second term vanishes because the vector basis is covariantly constant. Since $u_j$ is a scalar, its covariant derivative is
\bea
\nabla_\gamma\partial_\beta u_j=\partial_\gamma\partial_\beta u_j-\Gamma^\lambda_{\gamma\beta}\partial_\lambda u_j.
\eea
Then,
\bea
\nabla_\gamma\nabla_\beta\Psi_a^\mu=e_a^\mu\left(\partial_\gamma\partial_\beta u_j-\Gamma^\lambda_{\gamma\beta}\partial_\lambda u_j\right).
\eea
Contracting the derivative indices with the metric gives
\bea
\nabla_\beta\nabla^\beta\Psi_a^\mu&=&g^{\beta\gamma}\nabla_\gamma\nabla_\beta\Psi_a^\mu\nn\\
&=&e_a^\mu g^{\beta\gamma}\left(\partial_\gamma\partial_\beta u_j-\Gamma^\lambda_{\gamma\beta}\partial_\lambda u_j\right).
\eea
The expression in parentheses is precisely the scalar d'Alembertian acting on $u_j$,
\bea
\Box u_j=g^{\beta\gamma}\left(\partial_\gamma\partial_\beta u_j-\Gamma^\lambda_{\gamma\beta}\partial_\lambda u_j\right).
\eea
Hence, the vector Laplacian acting on $\Psi_a^\mu$ reduces to
\bea
\nabla_\beta\nabla^\beta\Psi_a^\mu=e_a^\mu\Box u_j.
\eea
Thus, whenever $u_j$ is an eigenfunction of the scalar d'Alembertian, the corresponding vector field $\Psi_a^\mu$ is an eigenfunction of the vector Laplacian with the same eigenvalue.

For a scalar mode with orbital angular momentum $l$, we write
\bea
u_l=J_l(|q_\perp|\rho)e^{-\ii E_l t}e^{\ii l\phi}e^{\ii q_zz},
\eea
where the on-shell condition following from the scalar wave equation is
\bea
(E_l+l\Omega)^2=|\mathbf{q}_\perp|^2+q_z^2.
\eea
It is useful to introduce the shifted frequency
\bea
\widetilde E_l=E_l+l\Omega,
\eea
so that, for positive-frequency modes,
\bea
\widetilde E_l=\sqrt{|\mathbf{q}_\perp|^2+q_z^2}.
\eea

The transverse vector basis is decomposed into circular components. Starting from the parallel tetrad basis, we define the circular tetrads as
\bea
e_\pm^\mu=\frac{1}{\sqrt{2}}\left(e_1^\mu\pm\ii e_2^\mu\right)=e^{\pm\ii\Omega t}\boldsymbol{\epsilon}_\pm^\mu,
\label{tetrads_and_epsolons}
\eea
where the constant polarization vectors are
\bea
\boldsymbol{\epsilon}_\pm^\mu=\frac{1}{\sqrt{2}}(0,1,\pm\ii,0).
\eea

The relation between the circular tetrads and the constant polarization vectors is important in identifying the common frequency of the vector modes. A scalar mode with orbital angular momentum $l$ combined with $e_\pm^\mu$ carries total angular momentum
\bea
j=l\pm1.
\eea
Therefore, for a fixed total angular momentum $j$, the transverse sector is generated by the scalar mode with $l=j-1$ in the $(+)$ sector and the scalar mode with $l=j+1$ in the $(-)$ sector. We denote by $E_j$ the temporal frequency of the complete vector modes in this fixed-$j$ sector. The corresponding shifted frequency is
\bea
\widetilde E_j=E_j+j\Omega.
\eea

Since the scalar modes entering the two transverse contributions are constructed with the same spatial momentum $\mathbf q$, their shifted energies are determined by the common momentum norm,
\bea
&&\widetilde E_{j-1}^2=|{\bf q}|^2=|{\bf q_\perp}|^2+q_z^2\nn\\
&&\widetilde E_j^2=|{\bf q}|^2=|{\bf q_\perp}|^2+q_z^2.
\eea
Choosing the positive-frequency branch, we therefore have
\bea
&&\widetilde E_{j-1}=|{\bf q}|=\widetilde E_j\nn\\
&\therefore&E_{j-1}+(j-1)\Omega=E_j+j\Omega
\eea
for the $(+)$ component. Hence,
\bea
E_{j-1}=E_j+\Omega.
\eea
The corresponding vector mode is therefore
\bea
\Psi_{+,j-1}^\mu&=&e_+^\mu u_{j-1}\nn\\
&=&e^{+\ii\Omega t}\boldsymbol{\epsilon}_+^\mu J_{j-1}(|q_\perp|\rho)e^{-\ii(E_j+\Omega)t}e^{\ii(j-1)\phi}e^{\ii q_zz}\nn\\
&=&J_{j-1}(|q_\perp|\rho)e^{-\ii E_jt}e^{\ii(j-1)\phi}e^{\ii q_zz}\boldsymbol{\epsilon}_+^\mu,
\label{Psi+}
\eea
where Eq.~\eqref{tetrads_and_epsolons} was used. 

For the $-$ component, the orbital angular momentum is $l=j+1$. Requiring the same shifted frequency gives
\bea
\widetilde E_{j+1}=E_{j+1}+(j+1)\Omega=\widetilde E_j=E_j+j\Omega,
\eea
and therefore
\bea
E_{j+1}=E_j-\Omega.
\eea
Using $e_-^\mu=e^{-\ii\Omega t}\boldsymbol{\epsilon}_-^\mu$, the corresponding mode becomes
\bea
\Psi_{-,j+1}^\mu&=&e_-^\mu u_{j+1}\nn\\
&=&e^{-\ii\Omega t}\boldsymbol{\epsilon}_-^\mu J_{j+1}(|q_\perp|\rho)e^{-\ii(E_j-\Omega)t}e^{\ii(j+1)\phi}e^{\ii q_zz}\nn\\
&=&J_{j+1}(|q_\perp|\rho)e^{-\ii E_jt}e^{\ii(j+1)\phi}e^{\ii q_zz}\boldsymbol{\epsilon}_-^\mu.
\label{Psi-}
\eea
Thus, although the scalar modes entering the two transverse contributions have different orbital angular momenta and different frequencies, the corresponding circular tetrads compensate these frequency shifts, so that both vector modes have the same temporal dependence $e^{-\ii E_jt}$. Their total angular momentum is the same,
\bea
J_z\Psi_{+,j-1}^\mu=j\Psi_{+,j-1}^\mu,\qquad J_z\Psi_{-,j+1}^\mu=j\Psi_{-,j+1}^\mu.
\eea

The longitudinal component is obtained from the scalar mode with orbital angular momentum $l=j$. Since the longitudinal tetrad $e_3^\mu$ is independent of $t$, $\rho$, and $\phi$, it does not contribute to the angular momentum or modify the temporal dependence. We therefore define
\bea
\Psi_{z,j}^\mu=e_3^\mu u_j=J_j(|q_\perp|\rho)e^{-\ii E_jt}e^{\ii j\phi}e^{\ii q_zz}\delta_z^\mu.
\label{Psiz}
\eea
This mode belongs to the same $j$ sector and has the same temporal frequency $E_j$. 

The temporal structure is constructed in the same way from the time-directed tetrad $e_0^\mu$. Since the tetrads are covariantly constant, the vector-Laplacian eigenfunctions are obtained by multiplying each tetrad by a scalar eigenfunction. Thus,
\bea
\Psi_{0,j}^\mu=e_0^\mu u_j=e_0^\mu J_j(|q_\perp|\rho)e^{-\ii E_jt}e^{\ii j\phi}e^{\ii q_zz}.
\label{Psi0}
\eea
Unlike the coordinate basis vector $\delta_0^\mu$, the tetrad $e_0^\mu$ is covariantly constant and therefore belongs to the vector-Laplacian eigenbasis constructed above. The four structures
\bea
\Psi_{0,j}^\mu,\qquad\Psi_{+,j-1}^\mu,\qquad\Psi_{-,j+1}^\mu,\qquad\Psi_{z,j}^\mu
\eea
provide the four independent vector-Laplacian eigenfunctions associated with a fixed total angular momentum $j$. The most general solution in this sector can therefore be written as
\bea
A_j^\mu=a_j\Psi_{0,j}^\mu+C_j^+\Psi_{+,j-1}^\mu+C_j^-\Psi_{-,j+1}^\mu+C_j^z\Psi_{z,j}^\mu.\nn\\
\eea
All four contributions have the same total angular momentum $j$, the same longitudinal momentum $q_z$, and the same temporal frequency $E_j$. Their different orbital angular momenta in the transverse sector are compensated by the corresponding angular dependence of the circular tetrads. This allows the divergence to be expressed in terms of the single scalar mode $u_j$,
\bea
\nabla_\mu A_j^\mu=\left[\frac{|q_\perp|}{\sqrt{2}}(C_j^--C_j^+)+\ii q_zC_j^z-\ii\widetilde E_ja_j\right]u_j.
\eea
The Lorenz condition therefore requires
\bea
\frac{|q_\perp|}{\sqrt{2}}(C_j^--C_j^+)+\ii q_zC_j^z-\ii\widetilde E_ja_j=0.
\eea
The gauge-fixed equation retains the residual gauge transformation
\bea
A^\mu\rightarrow A^\mu+\nabla^\mu\Lambda,
\eea
where the gauge function satisfies
\bea
\Box\Lambda=0.
\eea
Choosing $\Lambda$ with the same quantum numbers $(j,E_j,q_z,q_\perp)$, its gradient belongs to the same fixed-$j$ sector. For propagating modes, $\widetilde E_j\neq0$, and the residual gauge freedom can therefore be used to set
\bea
a_j=0.
\eea
The Lorenz condition then reduces to
\bea
\frac{|q_\perp|}{\sqrt{2}}(C_j^--C_j^+)+\ii q_zC_j^z=0.
\eea
The remaining coefficient space is two-dimensional. A convenient pair of independent solutions is obtained by choosing
\bea
(C_j^+,C_j^-,C_j^z)=(1,1,0),
\eea
and
\bea
(C_j^+,C_j^-,C_j^z)=(q_z,-q_z,-\ii\sqrt{2}|q_\perp|).
\eea
The corresponding physical photon modes are therefore
\bea
A_{j,1}^\mu=N_1\left[\Psi_{+,j-1}^\mu+\Psi_{-,j+1}^\mu\right],
\eea
and
\bea
A_{j,2}^\mu=N_2\left[q_z\Psi_{+,j-1}^\mu-q_z\Psi_{-,j+1}^\mu-\ii\sqrt{2}|q_\perp|\Psi_{z,j}^\mu\right].
\eea
Substituting the explicit expressions for the vector modes gives
\begin{widetext}
  \bea
A_{j,1}^\mu=N_1\left[J_{j-1}(|q_\perp|\rho)e^{\ii(j-1)\phi}\boldsymbol{\epsilon}_+^\mu+J_{j+1}(|q_\perp|\rho)e^{\ii(j+1)\phi}\boldsymbol{\epsilon}_-^\mu\right]e^{-\ii E_jt}e^{\ii q_zz},
\eea
and
\bea
A_{j,2}^\mu=N_2\left[q_zJ_{j-1}(|q_\perp|\rho)e^{\ii(j-1)\phi}\boldsymbol{\epsilon}_+^\mu-q_zJ_{j+1}(|q_\perp|\rho)e^{\ii(j+1)\phi}\boldsymbol{\epsilon}_-^\mu-\ii\sqrt{2}|q_\perp|J_j(|q_\perp|\rho)e^{\ii j\phi}\delta_z^\mu\right]e^{-\ii E_jt}e^{\ii q_zz}.
\eea  
\end{widetext}
}

\section{Normalization of the Physical Photon Modes}
\label{subsec:PhotonModeNormalization}
The physical photon modes must be normalized in order to define the
corresponding creation and annihilation operators in the field expansion.
For this purpose, we need an inner product on the space of physical
solutions of the Maxwell equations. Unlike the usual $L^2$ inner product
of nonrelativistic wave functions, the natural inner product for a
relativistic gauge field is constructed from a conserved current.

For two physical solutions $A^\mu$ and $B^\mu$ satisfying
Maxwell's equations,
\bea
&&\nabla_\mu F_A^{\mu\nu}=0,\qquad\nabla_\mu F_B^{\mu\nu}=0,
\eea
consider the bilinear current
\bea
&&j^\mu[A,B]=A_\nu^*F_B^{\mu\nu}-F_A^{*\mu\nu}B_\nu.
\label{PhotonMaxwellCurrent}
\eea
Using the antisymmetry of the field strength and the Maxwell equations,
this current is conserved,
\bea
&&\nabla_\mu j^\mu[A,B]=0.
\label{PhotonInnerProductCurrentConservation}
\eea

The conservation law implies that the flux of $j^\mu$ through any two
spacelike hypersurfaces $\Sigma_1$ and $\Sigma_2$ is the same, provided
the contribution from the boundary at spatial infinity vanishes. This
follows from the covariant Gauss theorem,
\bea
&&\int_{\Sigma_2}d\Sigma_\mu\,j^\mu-\int_{\Sigma_1}d\Sigma_\mu\,j^\mu
=\int_{\mathcal V}d^4x\,\sqrt{-g}\,\nabla_\mu j^\mu=0.\nn\\
\label{PhotonCurrentGaussTheorem}
\eea
Therefore, the flux through a spacelike hypersurface provides a
hypersurface-independent bilinear form on the space of physical
solutions. We use it to define the Maxwell inner product as
\bea
&& (A,B)=-\ii\int_\Sigma d\Sigma_\mu\left[A_\nu^*F_B^{\mu\nu}-F_A^{*\mu\nu}B_\nu\right].
\label{PhotonMaxwellInnerProduct}
\eea
The hypersurface $\Sigma$ specifies the spatial slice on which
the norm is evaluated. Because the associated current is conserved, the
result does not depend on which spacelike slice is chosen. We can
therefore choose the constant-$t$ hypersurface of the rotating
coordinates, which is convenient for evaluating the mode normalization.

For the rotating geometry considered here, $\sqrt{-g}=1$ and
$g^{00}=1$. Therefore, on a constant-$t$ hypersurface, the covariant
surface element is
\bea
&&d\Sigma_\mu=(1,0,0,0)d^3x.
\label{PhotonConstantTimeSurface}
\eea
The inner product can thus be evaluated as
\bea
&& (A,B)=-\ii\int d^3x\left[A_\nu^*F_B^{0\nu}-F_A^{*0\nu}B_\nu\right].
\label{PhotonConstantTimeInnerProduct}
\eea
Although the expression in Eq.~\eqref{PhotonConstantTimeInnerProduct}
has the same coordinate form as in Minkowski space, the quantities
$A_\nu$, $B_\nu$, $F^{0\nu}_A$, and $F^{0\nu}_B$ must be constructed using
the rotating metric. In particular, one must not set $A_0=0$ merely from
the choice $A^0=0$, since lowering the index with the rotating metric
gives
\bea
&&A_0=g_{0\nu}A^\nu.
\label{PhotonRotatingCovariantTimeComponent}
\eea

The normalization can nevertheless be evaluated in a simpler form by
using the fact that the rotating coordinates are related to inertial
coordinates by a coordinate transformation. In cylindrical coordinates,
the transformation is given by Eq.~\eqref{PhotonCoordinateTransformation}. Therefore, the hypersurface $t=\mathrm{const.}$ coincides with the hypersurface
$t'=\mathrm{const.}$, and the Maxwell inner product is invariant under
this coordinate transformation.

The physical modes constructed above have the common total angular
momentum eigenvalue $l$, as specified by the construction leading to
Eq.~\eqref{PhotonPhysicalConstraint}. Their common temporal dependence
is $e^{-\ii Et}$ and their azimuthal dependence belongs to the same
total-$J_z$ sector. Under Eq.~\eqref{PhotonCoordinateTransformation},
\bea
&&e^{-\ii Et}e^{\ii l\phi}=e^{-\ii(E+\Omega l)t'}e^{\ii l\phi'}.
\label{PhotonModeInertialTimeDependence}
\eea
The frequency measured with respect to inertial time is therefore
\bea
&&\widetilde E_l=E+\Omega l.
\label{PhotonNormalizationShiftedEnergy}
\eea
which is precisely the shifted energy introduced in
Eq.~\eqref{PhotonShiftedEnergy}. Using the dispersion relation already
obtained in Eq.~\eqref{PhotonMasslessDispersion}, this frequency satisfies
\bea
&&\widetilde E_l=\sqrt{|\mathbf q_\perp|^2+q_z^2}.
\label{PhotonNormalizationDispersion}
\eea
Thus, the normalization of a positive-frequency rotating mode is
controlled by $\widetilde E_l$, rather than directly by its rotating-frame
frequency $E$.

We now evaluate the spatial overlaps entering the Maxwell inner product
in Eq.~\eqref{PhotonConstantTimeInnerProduct}. For the physical modes
constructed above, the residual gauge freedom has been used to impose
$A^0=0$. Notice, however, that $A_0$ does not vanish in general in the
rotating coordinates, since $g_{0i}\neq0$. For a mode with total angular
momentum $l$, the time and angular dependence of the field gives
\bea
&&F^{0i}=-\ii\widetilde E_l A^i,
\label{PhotonTimeSpaceFieldStrength}
\eea
and the Maxwell inner product Then reduces to
\bea
&&(A,B)=2\widetilde E_l\int d^3x\,\mathbf A^*\cdot\mathbf B.
\label{PhotonReducedInnerProduct}
\eea
The two physical modes will be normalized in the symmetric basis
introduced in Eqs.~\eqref{PhotonSymmetricPolarizationBasis} and~\eqref{PhotonSymmetricPhysicalModes}.
In this basis, both modes contain the same three spatial structures, and
their dependence on the polarization sector is entirely contained in the
coefficients $\widetilde C_\beta^+$, $\widetilde C_\beta^-$, and
$\widetilde C_\beta^z$. We therefore consider the
spatial overlap between two arbitrary modes labeled by $\beta$ and
$\gamma$.

The cylindrical Bessel functions satisfy
\bea
&&\int_0^\infty\rho\,d\rho\,J_n(|\mathbf q_\perp|\rho)J_n(|\mathbf q_\perp'|\rho)=\frac{1}{|\mathbf q_\perp|}\delta(|\mathbf q_\perp|-|\mathbf q_\perp'|).\nn\\
\label{PhotonBesselModeNormalization}
\eea
The angular and longitudinal integrations give
\bea
&&\int_0^{2\pi}d\phi\,e^{\ii(l-l')\phi}=2\pi\delta_{ll'},
\label{PhotonAngularModeNormalization}
\eea
and
\bea
&&\int_{-\infty}^{\infty}dz\,e^{\ii(q_z-q_z')z}=2\pi\delta(q_z-q_z').
\label{PhotonLongitudinalModeNormalization}
\eea

The circular polarization vectors are orthonormal in the transverse
spatial subspace,
\bea
&&\boldsymbol{\epsilon}_\pm^*\cdot\boldsymbol{\epsilon}_\pm=1,\qquad\boldsymbol{\epsilon}_+^*\cdot\boldsymbol{\epsilon}_-=0,
\label{PhotonCircularSpatialOrthogonality}
\eea
and are orthogonal to the longitudinal direction,
\bea
&&\boldsymbol{\epsilon}_\pm^*\cdot\hat{\boldsymbol z}=0.
\label{PhotonCircularLongitudinalOrthogonality}
\eea

Using the common form of the physical modes in Eq.~\eqref{PhotonSymmetricPhysicalModes}, the spatial overlap between two arbitrary polarization sectors is therefore
{\small
\bea
\int d^3x\,\widetilde{\mathbf A}_{l,\beta}^*\cdot\widetilde{\mathbf A}_{l',\gamma}&=&\frac{4\pi^2\widetilde{\mathcal N}_\beta^*\widetilde{\mathcal N}_\gamma}{|\mathbf q_\perp|}\delta_{ll'}\delta(|\mathbf q_\perp|-|\mathbf q_\perp'|)\delta(q_z-q_z')\nn\\
&\times&\left(\widetilde C_\beta^{+*}\widetilde C_\gamma^++\widetilde C_\beta^{-*}\widetilde C_\gamma^-+\widetilde C_\beta^{z*}\widetilde C_\gamma^z\right).
\label{PhotonSymmetricSpatialOverlapGeneral}
\eea
}
The diagonal coefficient products follow directly from the definitions
of the symmetric basis. Note that for each polarization sector,
\bea
&&|\widetilde C_\beta^+|^2+|\widetilde C_\beta^-|^2+|\widetilde C_\beta^z|^2=1,
\label{PhotonSymmetricCoefficientNormOne}
\eea
and the orthogonality between the two polarization sectors follows from the
corresponding cross products. For the two transverse components, we find
\bea
&&\widetilde C_1^{+*}\widetilde C_2^+=\frac{1}{4}\left(1-\frac{q_z^2}{|\mathbf q|^2}\right),
\label{PhotonSymmetricCrossCoefficientPlus}
\eea
and
\bea
&&\widetilde C_1^{-*}\widetilde C_2^-=\frac{1}{4}\left(1-\frac{q_z^2}{|\mathbf q|^2}\right),
\label{PhotonSymmetricCrossCoefficientMinus}
\eea
whereas the longitudinal contribution is
\bea
&&\widetilde C_1^{z*}\widetilde C_2^z=-\frac{1}{2}\frac{|\mathbf q_\perp|^2}{|\mathbf q|^2}.
\label{PhotonSymmetricCrossCoefficientLongitudinal}
\eea

Using $|\mathbf q|^2=|\mathbf q_\perp|^2+q_z^2$, the sum of the three
contributions vanishes,
\bea
&&\widetilde C_1^{+*}\widetilde C_2^++\widetilde C_1^{-*}\widetilde C_2^-+\widetilde C_1^{z*}\widetilde C_2^z=0.
\label{PhotonSymmetricCrossCoefficientSum}
\eea
The diagonal and off-diagonal relations can therefore be summarized as
\bea
&&\sum_{s=+,-,z}\widetilde C_\beta^{s*}\widetilde C_\gamma^s=\delta_{\beta\gamma}.
\label{PhotonSymmetricCoefficientOrthogonality}
\eea
Substituting this relation into Eq.~\eqref{PhotonSymmetricSpatialOverlapGeneral},
we obtain
\bea
&&\int d^3x\,\widetilde{\mathbf A}_{l,\beta}^*\cdot\widetilde{\mathbf A}_{l',\gamma}\nn\\
&=&\frac{4\pi^2\widetilde{\mathcal N}_\beta^*\widetilde{\mathcal N}_\gamma}{|\mathbf q_\perp|}\delta_{ll'}\delta_{\beta\gamma}\delta(|\mathbf q_\perp|-|\mathbf q_\perp'|)\delta(q_z-q_z').\nn\\
\label{PhotonSymmetricSpatialOverlap}
\eea
Using the reduced Maxwell inner product in Eq.~\eqref{PhotonReducedInnerProduct},
the corresponding overlap of the physical photon modes is
\bea
&&( \widetilde A_{l,\beta},\widetilde A_{l',\gamma})\nn\\
&=&\frac{8\pi^2\widetilde E_l\widetilde{\mathcal N}_\beta^*\widetilde{\mathcal N}_\gamma}{|\mathbf q_\perp|}\delta_{ll'}\delta_{\beta\gamma}\delta(|\mathbf q_\perp|-|\mathbf q_\perp'|)\delta(q_z-q_z').\nn\\
\label{PhotonSymmetricMaxwellOverlap}
\eea
We normalize the physical modes according to the cylindrical convention
adopted in the previous normalization, namely,
\bea
&&( \widetilde A_{l,\beta}(\mathbf q),\widetilde A_{l',\gamma}(\mathbf q'))\nn\\
&=&\frac{1}{|\mathbf q_\perp|}\delta_{ll'}\delta_{\beta\gamma}\delta(|\mathbf q_\perp|-|\mathbf q_\perp'|)\delta(q_z-q_z').
\label{PhotonSymmetricNormalizationCondition}
\eea
which fixes
\bea
&&\widetilde{\mathcal N}_\beta^*\widetilde{\mathcal N}_\gamma=\frac{1}{8\pi^2\widetilde E_l}\delta_{\beta\gamma}.
\label{PhotonSymmetricNormalizationSquared}
\eea

Choosing the normalization constants real and positive gives the common
normalization
\bea
\widetilde{\mathcal N}_1=\widetilde{\mathcal N}_2=\frac{1}{2\pi(2\widetilde E_l)^{1/2}}.
\label{PhotonSymmetricNormalization}
\eea

\section{Field Expansion and Quantization of the Physical Photon Modes}
\label{subsec:PhotonFieldExpansion}

The physical photon modes form the basis for the quantization of the electromagnetic field. We therefore expand the field operator in terms of the two independent physical solutions given in Eqs.~\eqref{PhotonModeOneFinal} and \eqref{PhotonModeTwoFinal} and their complex conjugates. The mode functions are expressed in rotating coordinates with the temporal dependence $e^{-\ii Et}$. The corresponding inertial frequency is related to the rotating-frame frequency through $\widetilde E_l=E+\Omega l$, while the dispersion relation is given by $\widetilde E_l=\sqrt{|\mathbf q_\perp|^2+q_z^2}$. Thus, $\widetilde E_l$ enters the normalization through the Maxwell inner product, whereas $E$ remains the frequency parameter appearing explicitly in the rotating-frame modes.

The two physical polarization sectors are labelled by $\alpha=1,2$, while $l\in\mathbb Z$ denotes the total angular momentum sector. The continuous momentum variables are $|\mathbf q_\perp|$ and $q_z$, and the corresponding spectral measure is
\bea
&&\sumint_{l,\alpha}[dq]\equiv\sum_{l=-\infty}^{\infty}\sum_{\alpha=1,2}\int_{-\infty}^{\infty}dq_z\int_0^\infty|\mathbf q_\perp|\,d|\mathbf q_\perp|.\nn\\
\label{PhotonModeMeasure}
\eea
The factor $|\mathbf q_\perp|$ is the radial measure associated with the cylindrical momentum representation. It is consistent with the normalization of the Bessel functions in Eq.~\eqref{PhotonBesselModeNormalization} and with the orthonormality relation in Eq.~\eqref{PhotonSymmetricMaxwellOverlap}. No additional factors of $2\pi$ are included in the measure, since the factors generated by the angular and longitudinal integrations are already incorporated into the normalization constants $\mathcal N_1$ and $\mathcal N_2$.

The quantized electromagnetic potential is then expanded in the physical mode basis as
\bea
&&\hat A^\mu(x)=\sumint_{l,\alpha}[dq]\left[\hat a_{l,\alpha}(\mathbf q)A_{l,\alpha}^\mu(x)+\hat a_{l,\alpha}^\dagger(\mathbf q)A_{l,\alpha}^{\mu *}(x)\right].\nn\\
\label{PhotonFieldExpansion}
\eea
The first term contains the independent physical modes, while the second contains their complex conjugates. The pairing of each mode with its complex conjugate ensures the Hermiticity of the electromagnetic field operator. Since the expansion is restricted to the two physical polarization sectors, no additional scalar or gauge modes are included.
Because the classical modes satisfy the source-free Maxwell equations, the linear combination appearing in the operator expansion satisfies the corresponding operator equation. 

The mode coefficients can be extracted from the field operator using the Maxwell inner product defined in Eq.~\eqref{PhotonMaxwellInnerProduct}. In particular, it is straightforward to show that
\bea
&&\sumint_{l',\beta}[dq']\hat a_{l',\beta}(\mathbf q')(A_{l,\alpha}(\mathbf q),A_{l',\beta}(\mathbf q'))=\hat a_{l,\alpha}(\mathbf q).\nn\\
\label{PhotonAnnihilationProjectionResult}
\eea

We now consider the mixed inner product $(A_{l,\alpha}(\mathbf q),A_{l',\beta}^*(\mathbf q'))$ appearing in the field projection. Taking $A=A_{l,\alpha}$ and $B=A_{l',\beta}^*$ in the Maxwell inner product defined in Eq.~\eqref{PhotonMaxwellInnerProduct}, and using the reduction to the spatial components established in Eq.~\eqref{PhotonReducedInnerProduct}, we obtain
\bea
&&(A_{l,\alpha}(\mathbf q),A_{l',\beta}^*(\mathbf q'))\nn\\
&=&-\ii\int d^3x\left[A_{l,\alpha}^{*i}F_{A_{l',\beta}^*}^{0i}-F_{A_{l,\alpha}}^{*0i}A_{l',\beta}^{*i}\right].
\label{PhotonMixedInnerProduct}
\eea
Using Eq.~\eqref{PhotonTimeSpaceFieldStrength} and its complex conjugate, we have
\bea
F_{A_{l',\beta}^*}^{0i}&=&+\ii\widetilde E_{l'}A_{l',\beta}^{*i},\nn\\
F_{A_{l,\alpha}}^{*0i}&=&+\ii\widetilde E_lA_{l,\alpha}^{*i}.
\eea
Therefore,
\bea
&&A_{l,\alpha}^{*i}F_{A_{l',\beta}^*}^{0i}-F_{A_{l,\alpha}}^{*0i}A_{l',\beta}^{*i}\nn\\
&&=\ii\widetilde E_{l'}A_{l,\alpha}^{*i}A_{l',\beta}^{*i}-\ii\widetilde E_lA_{l,\alpha}^{*i}A_{l',\beta}^{*i}\nn\\
&&=\ii(\widetilde E_{l'}-\widetilde E_l)A_{l,\alpha}^{*i}A_{l',\beta}^{*i}.
\label{PhotonMixedInnerProductIntegrand}
\eea

The spatial overlap of the complex-conjugated modes is nonzero only for the corresponding conjugate momentum labels. In particular, the angular and longitudinal dependences require $l'=-l$ and $q_z'=-q_z$, while $|\mathbf q_\perp'|=|\mathbf q_\perp|$. Since the dispersion relation in Eq.~\eqref{PhotonNormalizationDispersion} depends only on $|\mathbf q_\perp|$ and $q_z^2$, these conditions give $\widetilde E_{l'}=\widetilde E_l$. Hence, the contribution in Eq.~\eqref{PhotonMixedInnerProductIntegrand} vanishes, and
\bea
&&(A_{l,\alpha}(\mathbf q),A_{l',\beta}^*(\mathbf q'))=0.
\label{PhotonPositiveNegativeOrthogonality1}
\eea

Similarly, the projection onto the complex-conjugated mode gives
\bea
&&(A_{l,\alpha}^*(\mathbf q),A_{l',\beta}^*(\mathbf q'))=-(A_{l',\beta}(\mathbf q'),A_{l,\alpha}(\mathbf q)),
\label{PhotonConjugateInnerProduct}
\eea
so that
\bea
&&(A_{l,\alpha}^*(\mathbf q),\hat A)=-\hat a_{l,\alpha}^\dagger(\mathbf q),
\label{PhotonCreationProjectionResult}
\eea
and
\bea
&&(A_{l,\alpha}(\mathbf q),A_{l',\beta}^*(\mathbf q'))=0.
\label{PhotonPositiveNegativeOrthogonality2}
\eea

Upon quantization, the coefficients of the physical photon modes are promoted to bosonic annihilation and creation operators. The bosonic commutation structure follows from canonical quantization of the electromagnetic field, while the normalization of the corresponding mode operators is fixed by the Maxwell inner product of the physical modes in Eq.~\eqref{PhotonSymmetricMaxwellOverlap} and by the mode measure defined in Eq.~\eqref{PhotonModeMeasure}. Thus, the canonical commutation relation takes the form
\bea
&&[\hat a_{l,\alpha}(\mathbf q),\hat a_{l',\beta}^\dagger(\mathbf q')]\nn\\
&=&\frac{1}{|\mathbf q_\perp|}\delta_{ll'}\delta_{\alpha\beta}\delta(|\mathbf q_\perp|-|\mathbf q_\perp'|)\delta(q_z-q_z').
\label{PhotonCommutationRelations}
\eea
The annihilation operators commute among themselves,
\bea
&&[\hat a_{l,\alpha}(\mathbf q),\hat a_{l',\beta}(\mathbf q')]=0,
\label{PhotonAnnihilationCommutationRelations}
\eea
and the creation operators likewise satisfy
\bea
&&[\hat a_{l,\alpha}^\dagger(\mathbf q),\hat a_{l',\beta}^\dagger(\mathbf q')]=0.
\label{PhotonCreationCommutationRelations}
\eea

The photon vacuum is defined as the state annihilated by all annihilation operators,
\bea
&&\hat a_{l,\alpha}(\mathbf q)|0_\Omega\rangle_\gamma=0.
\label{PhotonVacuum}
\eea
The physical photon states are then constructed by acting with the creation operators on the vacuum. A one-photon state with angular momentum label $l$, physical polarization $\alpha$, and momentum $\mathbf q$ is defined as
\bea
&&|l,\alpha,\mathbf q\rangle_\gamma=\hat a_{l,\alpha}^\dagger(\mathbf q)|0_\Omega\rangle_\gamma.
\label{PhotonOneParticleState}
\eea

More generally, multiphoton states are generated by successive applications of creation operators,
\bea
&&|l_1,\alpha_1,\mathbf q_1;\ldots;l_n,\alpha_n,\mathbf q_n\rangle_\gamma\nn\\
&=&\hat a_{l_1,\alpha_1}^\dagger(\mathbf q_1)\cdots\hat a_{l_n,\alpha_n}^\dagger(\mathbf q_n)|0_\Omega\rangle_\gamma.
\label{PhotonMultiparticleStates}
\eea

Since the creation operators commute, these states are symmetric under the exchange of any two photons, as required for a bosonic field. The commutation relation in Eq.~\eqref{PhotonCommutationRelations} determines the normalization and orthogonality of the one-photon states. In particular,
\bea
&&{}_\gamma\langle l,\alpha,\mathbf q|l',\beta,\mathbf q'\rangle_\gamma\nn\\
&=&\frac{1}{|\mathbf q_\perp|}\delta_{ll'}\delta_{\alpha\beta}\delta(|\mathbf q_\perp|-|\mathbf q_\perp'|)\delta(q_z-q_z'),
\label{PhotonOneParticleNormalization}
\eea
where the corresponding bra is
\bea
&&{}_\gamma\langle l,\alpha,\mathbf q|={}_\gamma\langle0_\Omega|\hat a_{l,\alpha}(\mathbf q).
\eea

\section{The Physical Photon Propagator}
\label{subsec:PhotonPropagator}

The quantized field defined in Eq.~\eqref{PhotonFieldExpansion} determines the photon propagator once the vacuum and the canonical commutation relations are specified. We define the Feynman propagator in the rotating coordinates by
\bea
&&D_\text{F}^{\mu\nu}(x,x')=\langle0_\Omega|T\left\{\hat A^\mu(x)\hat A^\nu(x')\right\}|0_\Omega\rangle_\gamma.
\label{PhotonFeynmanPropagator}
\eea
Since the time-ordering operation is defined with respect to the coordinate $t$, this propagator can be written as
\bea
&&D_\text{F}^{\mu\nu}(x,x')=\theta(t-t')D_>^{\mu\nu}(x,x')+\theta(t'-t)D_<^{\mu\nu}(x,x'),\nn\\
\label{PhotonTimeOrderedDecomposition}
\eea
where the two contributions to the time ordered two point function are obtained directly from the mode expansion.

For $t>t'$, the time ordering reduces to the ordinary operator product. Substituting the field expansion of Eq.~\eqref{PhotonFieldExpansion} into the two field operators gives
\bea
&&D_>^{\mu\nu}(x,x')\nn\\
&=&\sumint_{l,\alpha}[dq]\sumint_{l',\beta}[dq']\Big\langle0_\Omega\Big|\left[\hat a_{l,\alpha}(\mathbf q)A_{l,\alpha}^{\mu}(x)+\hat a_{l,\alpha}^{\dagger}(\mathbf q)A_{l,\alpha}^{\mu *}(x)\right]\nn\\
&&\times\left[\hat a_{l',\beta}(\mathbf q')A_{l',\beta}^{\nu}(x')+\hat a_{l',\beta}^{\dagger}(\mathbf q')A_{l',\beta}^{\nu *}(x')\right]\Big|0_\Omega\Big\rangle_\gamma.\nn\\
\label{PhotonGreaterExpansion}
\eea

Using Eq.~\eqref{PhotonVacuum}, the terms containing $\hat a_{l',\beta}(\mathbf q')$ on the right vanish. The remaining terms are
\bea
D_>^{\mu\nu}(x,x')&=&\sumint_{l,\alpha}[dq]\sumint_{l',\beta}[dq']A_{l,\alpha}^{\mu}(x)A_{l',\beta}^{\nu *}(x')\nn\\
&\times&\langle0_\Omega|\hat a_{l,\alpha}(\mathbf q)\hat a_{l',\beta}^{\dagger}(\mathbf q')|0_\Omega\rangle_\gamma.\nn\\
\label{PhotonGreaterVacuum}
\eea
Since $\hat a_{l,\alpha}(\mathbf q)|0_\Omega\rangle_\gamma=0$, the matrix element can be evaluated by commuting the annihilation operator through the creation operator:
\bea
&&\langle0_\Omega|\hat a_{l,\alpha}(\mathbf q)\hat a_{l',\beta}^{\dagger}(\mathbf q')|0_\Omega\rangle_\gamma=\langle0_\Omega|[\hat a_{l,\alpha}(\mathbf q),\hat a_{l',\beta}^{\dagger}(\mathbf q')]|0_\Omega\rangle_\gamma.\nn\\
\label{PhotonGreaterCommutator}
\eea
Using the canonical commutation relation in Eq.~\eqref{PhotonCommutationRelations}, we obtain
\bea
&&\langle0_\Omega|\hat a_{l,\alpha}(\mathbf q)\hat a_{l',\beta}^{\dagger}(\mathbf q')|0_\Omega\rangle_\gamma\nn\\
&=&\frac{1}{|\mathbf q_\perp|}\delta_{ll'}\delta_{\alpha\beta}\delta(|\mathbf q_\perp|-|\mathbf q_\perp'|)\delta(q_z-q_z').
\label{PhotonGreaterMatrixElement}
\eea
The integrations over the primed quantum numbers are then performed using the delta functions, yielding
\bea
&&D_>^{\mu\nu}(x,x')=\sumint_{l,\alpha}[dq]\,A_{l,\alpha}^{\mu}(x)A_{l,\alpha}^{\nu *}(x').
\label{PhotonGreaterWightman}
\eea
The second contribution is obtained by reversing the order of the two field operators,
\bea
&&D_<^{\mu\nu}(x,x')=\sumint_{l,\alpha}[dq]\,A_{l,\alpha}^{\mu *}(x)A_{l,\alpha}^{\nu}(x').
\label{PhotonLesserWightman}
\eea
For $t'<t$, this contribution enters the time ordered propagator in Eq.~\eqref{PhotonTimeOrderedDecomposition}. The mode labels appearing on the right-hand sides are summed and integrated according to the measure in Eq.~\eqref{PhotonModeMeasure}. Substitution of Eqs.~\eqref{PhotonGreaterWightman} and \eqref{PhotonLesserWightman} into Eq.~\eqref{PhotonTimeOrderedDecomposition} gives the physical-mode representation
\bea
D_\text{F}^{\mu\nu}(x,x')&=&\sumint_{l,\alpha}[dq]\Big[\theta(t-t')A_{l,\alpha}^{\mu}(x)A_{l,\alpha}^{\nu *}(x')\nn\\
&+&\theta(t'-t)A_{l,\alpha}^{\mu *}(x)A_{l,\alpha}^{\nu}(x')\Big].
\label{PhotonPhysicalPropagator}
\eea

To make the spatial and temporal structure of the propagator explicit, we write the physical photon modes in terms of their cylindrical components. For this purpose, define
\bea
&&\Phi_{l}(\mathbf x_\perp;\mathbf q_\perp)=J_l(|\mathbf q_\perp|\rho)e^{\ii l\phi}.
\label{PhotonCylindricalModeFunction}
\eea
The two physical modes introduced above can then be written in a common form,
\begin{widetext}
\bea
\widetilde A_{l,\beta}^\mu(x)=\widetilde{\mathcal N}_\beta\Big[\widetilde C_\beta^+\Phi_{l,l-1}(\mathbf x_\perp;\mathbf q_\perp)\boldsymbol{\epsilon}_+^\mu+\widetilde C_\beta^-\Phi_{l,l+1}(\mathbf x_\perp;\mathbf q_\perp)\boldsymbol{\epsilon}_-^\mu+\widetilde C_\beta^z\Phi_{l,l}(\mathbf x_\perp;\mathbf q_\perp)\delta_z^\mu\Big]e^{-\ii Et}e^{\ii q_z z},
\label{PhotonSymmetricModeSpatialStructure}
\eea
\end{widetext}
where $\beta=1,2$ labels the two independent physical polarization sectors. The coefficients $\widetilde C_\beta^+$, $\widetilde C_\beta^-$, and $\widetilde C_\beta^z$ contain the polarization dependence, while the three cylindrical structures are common to both modes.

Using Eq.~\eqref{PhotonCylindricalModeFunction}, the transverse part of Eq.~\eqref{PhotonSymmetricModeSpatialStructure} can be written as
\bea
&&\widetilde C_\beta^+\Phi_{l,l-1}\boldsymbol{\epsilon}_+^\mu+\widetilde C_\beta^-\Phi_{l,l+1}\boldsymbol{\epsilon}_-^\mu\nn\\
&=&e^{\ii l\phi}\Big[\widetilde C_\beta^+J_{l-1}(|\mathbf q_\perp|\rho)e^{-\ii\phi}\boldsymbol{\epsilon}_+^\mu+\widetilde C_\beta^-J_{l+1}(|\mathbf q_\perp|\rho)e^{\ii\phi}\boldsymbol{\epsilon}_-^\mu\Big].\nn\\
\label{PhotonSymmetricModeTransverseExpanded}
\eea
To express this result in the cylindrical basis, we use the unit vectors
\bea
\hat{\boldsymbol{\rho}}^\mu&=&(0,\cos\phi,\sin\phi,0)^T,\nn\\
\hat{\boldsymbol{\phi}}^\mu&=&(0,-\sin\phi,\cos\phi,0)^T,
\label{CylindricalUnitVectors}
\eea
together with the circular polarization relations
\bea
e^{-\ii\phi}\boldsymbol{\epsilon}_+^\mu&=&\frac{1}{\sqrt{2}}\left(\hat{\boldsymbol{\rho}}^\mu+\ii\hat{\boldsymbol{\phi}}^\mu\right),\nn\\
e^{\ii\phi}\boldsymbol{\epsilon}_-^\mu&=&\frac{1}{\sqrt{2}}\left(\hat{\boldsymbol{\rho}}^\mu-\ii\hat{\boldsymbol{\phi}}^\mu\right).
\label{CircularPolarizationCylindricalRelations}
\eea
Substitution into Eq.~\eqref{PhotonSymmetricModeTransverseExpanded} gives
\bea
&&\widetilde C_\beta^+\Phi_{l,l-1}\boldsymbol{\epsilon}_+^\mu+\widetilde C_\beta^-\Phi_{l,l+1}\boldsymbol{\epsilon}_-^\mu\nn\\
&=&\frac{e^{\ii l\phi}}{\sqrt{2}}\Big[\left(\widetilde C_\beta^+J_{l-1}(|\mathbf q_\perp|\rho)+\widetilde C_\beta^-J_{l+1}(|\mathbf q_\perp|\rho)\right)\hat{\boldsymbol{\rho}}^\mu\nn\\
&+&\ii\left(\widetilde C_\beta^+J_{l-1}(|\mathbf q_\perp|\rho)-\widetilde C_\beta^-J_{l+1}(|\mathbf q_\perp|\rho)\right)\hat{\boldsymbol{\phi}}^\mu\Big].
\label{PhotonSymmetricModeCylindricalComponents}
\eea
The longitudinal contribution is already expressed in the cylindrical basis,
\bea
&&\widetilde C_\beta^z\Phi_{l,l}\delta_z^\mu=\widetilde C_\beta^zJ_l(|\mathbf q_\perp|\rho)e^{\ii l\phi}\delta_z^\mu.
\label{PhotonSymmetricModeLongitudinalComponent}
\eea

The cylindrical representation of the spatial part of the physical modes is therefore
{\small
\bea
\widetilde{\mathcal U}_{l,\beta}^\mu(\mathbf x_\perp;\mathbf q)&\equiv&\frac{e^{\ii l\phi}}{\sqrt{2}}\Big[\left(\widetilde C_\beta^+J_{l-1}(|\mathbf q_\perp|\rho)+\widetilde C_\beta^-J_{l+1}(|\mathbf q_\perp|\rho)\right)\hat{\boldsymbol{\rho}}^\mu\nn\\
&+&\ii\left(\widetilde C_\beta^+J_{l-1}(|\mathbf q_\perp|\rho)-\widetilde C_\beta^-J_{l+1}(|\mathbf q_\perp|\rho)\right)\hat{\boldsymbol{\phi}}^\mu\nn\\
&+&\sqrt{2}\widetilde C_\beta^zJ_l(|\mathbf q_\perp|\rho)\delta_z^\mu\Big].
\label{PhotonSymmetricCylindricalSpatialStructure}
\eea
}
so that
\bea
\widetilde A_{l,\beta}^\mu(x)=\widetilde{\mathcal N}_\beta\widetilde{\mathcal U}_{l,\beta}^\mu(\mathbf x_\perp;\mathbf q)e^{-\ii Et}e^{\ii q_z z}.
\label{PhotonSymmetricModeCompact}
\eea

The propagator kernels can now be constructed directly from the common spatial structures. For each physical polarization sector, define
\bea
\mathcal P_{l,\beta}^{\mu\nu}(\mathbf x'_\perp,\mathbf x_\perp;\mathbf q)\equiv|\widetilde{\mathcal N}_\beta|^2\widetilde{\mathcal U}_{l,\beta}^\mu(\mathbf x_\perp;\mathbf q)\,\widetilde{\mathcal U}_{l,\beta}^{\nu *}(\mathbf x'_\perp;\mathbf q).
\label{PhotonSymmetricPropagatorKernel}
\eea
The total physical polarization kernel is obtained by summing over the two independent physical sectors,
\bea
\mathcal P_l^{\mu\nu}(\mathbf x'_\perp,\mathbf x_\perp;\mathbf q)\equiv\sum_{\beta=1}^{2}\mathcal P_{l,\beta}^{\mu\nu}(\mathbf x'_\perp,\mathbf x_\perp;\mathbf q).
\label{PhotonPhysicalPolarizationKernelCompact}
\eea
Equivalently, using the common normalization obtained above,
\bea
\mathcal P_l^{\mu\nu}(\mathbf x'_\perp,\mathbf x_\perp;\mathbf q)=|\widetilde{\mathcal N}|^2\sum_{\beta=1}^{2}\widetilde{\mathcal U}_{l,\beta}^\mu(\mathbf x_\perp;\mathbf q)\widetilde{\mathcal U}_{l,\beta}^{\nu *}(\mathbf x'_\perp;\mathbf q),\nn\\
\label{PhotonPhysicalPolarizationKernelSymmetric}
\eea

The greater and lesser components of the physical photon propagator follow directly from the mode expansion. The greater function is
\bea
D_>^{\mu\nu}(x,x')=\sumint_l[dq]\,e^{-\ii E(t-t')}e^{\ii q_z(z-z')}\mathcal P_l^{\mu\nu}(\mathbf x'_\perp,\mathbf x_\perp;\mathbf q),\nn\\
\label{PhotonGreaterModeRepresentation}
\eea
whereas the lesser function is
\bea
D_<^{\mu\nu}(x,x')=\sumint_l[dq]\,e^{\ii E(t-t')}e^{-\ii q_z(z-z')}\mathcal P_l^{\nu\mu}(\mathbf x'_\perp,\mathbf x_\perp;\mathbf q).\nn\\
\label{PhotonLesserModeRepresentation}
\eea
where
\bea
\sumint_l[dq]\equiv\sum_{l=-\infty}^{\infty}\int_{-\infty}^{\infty}dq_z\int_0^\infty|\mathbf q_\perp|\,d|\mathbf q_\perp|.
\label{PhotonReducedModeMeasure}
\eea

Moreover, by defining $\Delta t=t-t'$ and using the on-shell relation $\widetilde E_l=E+l\Omega=|\mathbf q|$, the two time-ordering factors can be represented spectrally as
\begin{subequations}
\bea
\theta(\Delta t)e^{-\ii E_l\Delta t}=\int_{-\infty}^{+\infty}\frac{dE}{2\pi}\frac{\ii e^{-\ii E\Delta t}}{E-E_l+\ii\epsilon},
\eea
and
\bea
\theta(-\Delta t)e^{\ii E_l\Delta t}=-\int_{-\infty}^{+\infty}\frac{dE}{2\pi}\frac{\ii e^{-\ii E\Delta t}}{E+E_l-\ii\epsilon}.
\eea
\end{subequations}
Since the inertial frequency of a mode with total angular momentum $l$ is shifted according to $E_l=|\mathbf q|-l\Omega$, the time-ordered physical photon propagator takes the spectral form
\begin{widetext}
\bea
D_\text{F}^{\mu\nu}(x,x')=\ii\sumint_l[dq]\int_{-\infty}^{+\infty}\frac{dE}{2\pi}e^{-\ii E(t-t')}\left[\frac{e^{\ii q_z(z-z')}\mathcal P_l^{\mu\nu}(\mathbf x'_\perp,\mathbf x_\perp;\mathbf q)}{E-|\mathbf q|+l\Omega+\ii\epsilon}-\frac{e^{-\ii q_z(z-z')}\mathcal P_l^{\nu\mu}(\mathbf x'_\perp,\mathbf x_\perp;\mathbf q)}{E+|\mathbf q|-l\Omega-\ii\epsilon}\right].
\label{PhotonPhysicalSpectralPropagatorSymmetric}
\eea
\end{widetext}

The expression above provides the physical photon propagator in the rotating frame in a spectral representation adapted to the symmetric polarization basis. The angular momentum dependence enters through the shifted spectral poles, while the polarization kernel retains the explicit transverse spatial dependence of the physical modes. The two independent physical polarization sectors are summed only after constructing their individual mode contributions, so that the resulting propagator contains the complete physical polarization structure without introducing additional photon degrees of freedom. The representation separates the temporal spectral dependence, longitudinal propagation, and transverse cylindrical structure while retaining the angular momentum dependence associated with the rotating frame.

{

\section{THE FULL GAUGE-FIXED PHOTON PROPAGATOR}
\label{sec:FullGaugeFixedPhotonPropagator}

\subsection{Gauge-fixed propagator from the scalar Green function}\label{secIXA}

The physical photon propagator constructed in Sec.~\ref{subsec:PhotonPropagator} is obtained from the two independent physical polarization modes of the quantized electromagnetic field. It is therefore adapted to correlation functions involving physical photon states.

In a perturbative application of QED, however, internal photon lines require the Green function associated with the complete gauge-fixed Maxwell operator. In this section, we construct this propagator in the rotating coordinates and verify that it satisfies the corresponding Green-function equation.

In the construction below, it is important to distinguish the total angular momentum quantum number $j$ that labels the complete photon mode from the orbital angular momentum quantum number $l$ associated with an individual field component. As established in Sec.~\ref{subsec:PhotonModes}, the physical vector modes are organized into sectors of fixed total angular momentum,
\bea
\hat J_z A^\mu_j=jA^\mu_j.
\eea
The different components of a given mode can nevertheless carry different orbital angular momenta. In particular, the temporal and longitudinal components have $l=j$, whereas the circular components have $l=j-1$ and $l=j+1$, respectively. Thus, $j$ is the common quantum number of the complete physical state, while $l$ characterizes the orbital angular dependence of a particular component.

The full gauge-fixed propagator, however, is most conveniently constructed through an auxiliary scalar Green function. Since this scalar object carries no spin angular momentum, its orbital angular momentum coincides with the total angular momentum of the corresponding sector. We therefore use $j$ as the angular momentum label of the scalar cylindrical modes in this section. This convention should not be confused with the orbital angular momentum of the individual circular components of the vector photon modes.

In the Feynman gauge, $\xi=1$, the gauge-fixed field equation obtained in Sec.~\ref{subsec:PhotonWaveEquation} is given in Eq.~\eqref{PhotonFeynmanGaugeWaveEquation}
\bea
\nabla_\beta\nabla^\beta A^\mu=0,
\label{PhotonFullGaugeFieldEquation}
\eea
so that for the rotating coordinates considered here, the spacetime is related to Minkowski spacetime by a coordinate transformation. Moreover, the Ricci tensor therefore vanishes, as established in Eq.~\eqref{PhotonRicciZero}. Nevertheless, the covariant derivatives acting on the vector field contain the nontrivial Christoffel connections associated with the rotating coordinates.

We therefore distinguish the covariant vector operator from the scalar d'Alembertian and define
\bea
\mathcal K^\mu{}_\alpha\equiv\delta^\mu_\alpha\nabla_\beta\nabla^\beta.
\label{PhotonFullGaugeOperator}
\eea

The full gauge-fixed photon propagator is then defined through
\bea
\mathcal K^\mu{}_\alpha(x)D^{\alpha\nu}(x,x')=-\ii g^{\mu\nu}(x)\delta^{(4)}(x-x').
\label{PhotonFullGaugePropagatorDefinition}
\eea

A convenient construction follows from the scalar Green function in the rotating coordinates. The scalar d'Alembertian is defined in Eq.~\eqref{PhotonScalarDAlembertianDefinition} and, for the rotating metric, its cylindrical representation is given in Eq.~\eqref{PhotonDAlembertianCylindrical}
\bea
\Box=\left(\partial_t-\Omega\partial_\phi\right)^2-\partial_\rho^2-\frac{1}{\rho}\partial_\rho-\frac{1}{\rho^2}\partial_\phi^2-\partial_z^2.
\label{PhotonFullGaugeScalarDAlembertian}
\eea

We consider the cylindrical scalar modes introduced in Eq.~\eqref{PhotonCylindricalModeFunction}. Since the scalar field carries no intrinsic angular momentum, the orbital angular momentum of these modes is also their total angular momentum. We therefore label them by $j$,
\bea
u_j(x;q)=J_j(|\mathbf q_\perp|\rho)e^{\ii j\phi}e^{\ii q_z z}e^{-\ii Et}.
\label{PhotonFullGaugeScalarModes}
\eea
For this scalar sector, one therefore has $l=j$.

This identification is specific to the scalar Green function. It should not be confused with the orbital quantum number of the circular components of the full vector photon modes, for which $l=j\mp1$.

The action of the rotational derivative on these modes gives
\bea
(\partial_t-\Omega\partial_\phi)u_j(x;q)=-\ii(E+j\Omega)u_j(x;q),
\label{PhotonFullGaugeRotationalDerivative}
\eea
while the radial, angular, and longitudinal derivatives yield
\bea
\left[\partial_\rho^2+\frac{1}{\rho}\partial_\rho-\frac{j^2}{\rho^2}\right]J_j(|\mathbf q_\perp|\rho)=-|\mathbf q_\perp|^2J_j(|\mathbf q_\perp|\rho),
\label{PhotonFullGaugeBesselEquation}
\eea
together with $\partial_\phi^2e^{\ii j\phi}=-j^2e^{\ii j\phi}$ and $\partial_z^2e^{\ii q_z z}=-q_z^2e^{\ii q_z z}$ we obtain
\bea
\Box u_j(x;q)=-\left[(E+j\Omega)^2-|\mathbf q|^2\right]u_j(x;q),
\label{PhotonFullGaugeScalarEigenvalue}
\eea
where
\bea
|\mathbf q|^2=|\mathbf q_\perp|^2+q_z^2.
\eea
Thus, the cylindrical scalar modes are eigenfunctions of the scalar d'Alembertian with eigenvalue
\bea
\lambda_j(E,\mathbf q)=-\left[(E+j\Omega)^2-|\mathbf q|^2\right].
\label{PhotonFullGaugeScalarEigenvalueDefinition}
\eea

The Green function associated with the scalar d'Alembertian is defined by
\bea
\Box G(x,x')=-\ii\delta^{(4)}(x-x').
\label{PhotonFullGaugeScalarGreenEquation}
\eea

In order to construct $G(x,x')$, recall that for a differential operator $\mathcal O$ whose eigenfunctions $\psi_a(x)$ satisfy
\bea
\mathcal O\psi_a(x)=\lambda_a\psi_a(x),
\label{PhotonFullGaugeGeneralEigenvalueEquation}
\eea
the corresponding Green function can be expanded in the same eigenfunction basis. The spectral coefficient is obtained by inverting the eigenvalue of the operator, so that
\bea
G(x,x')=\sum_a\frac{-\ii}{\lambda_a-\ii\epsilon}\psi_a(x)\psi_a^*(x').
\label{PhotonFullGaugeGeneralSpectralRepresentation}
\eea
where the sum over $a$ represents the complete set of discrete and continuous quantum numbers of the eigenfunctions. For continuous labels, the corresponding sums are replaced by integrals with the appropriate spectral measure.

For the cylindrical eigenfunctions considered here, the discrete angular momentum label is $j$, while $|\mathbf q_\perp|$, $q_z$, and $E$ are continuous variables. Since the scalar modes carry no spin angular momentum, this discrete label is simultaneously their orbital and total angular momentum. The completeness relation in the transverse plane is therefore
\bea
&&\frac{1}{2\pi}\sum_{j=-\infty}^{+\infty}\int_0^\infty |\mathbf q_\perp|d|\mathbf q_\perp|e^{\ii j(\phi-\phi')}J_j(|\mathbf q_\perp|\rho)J_j(|\mathbf q_\perp|\rho')
\nn\\
&=&\frac{\delta(\rho-\rho')}{\rho}\delta(\phi-\phi'),
\label{PhotonFullGaugeCylindricalCompleteness}
\eea
while the temporal and longitudinal coordinates are represented through
\bea
\delta(t-t')&=&\int_{-\infty}^{+\infty}\frac{dE}{2\pi}e^{-\ii E(t-t')},
\nn\\
\delta(z-z')&=&\int_{-\infty}^{+\infty}\frac{dq_z}{2\pi}e^{\ii q_z(z-z')}.
\label{PhotonFullGaugeFourierCompleteness}
\eea

Substituting the cylindrical modes and their eigenvalue into the general spectral representation then gives
\begin{widetext}
  \bea
G(x,x')=\frac{\ii}{2\pi}\sum_{j=-\infty}^{+\infty}\int_0^\infty |\mathbf q_\perp|d|\mathbf q_\perp|\int_{-\infty}^{+\infty}\frac{dq_z}{2\pi}\int_{-\infty}^{+\infty}\frac{dE}{2\pi}\frac{ e^{-\ii E(t-t')}e^{\ii j(\phi-\phi')}e^{\ii q_z(z-z')}J_j(|\mathbf q_\perp|\rho)J_j(|\mathbf q_\perp|\rho')}{(E+j\Omega)^2-|\mathbf q|^2+\ii\epsilon}.
\label{PhotonFullGaugeScalarGreenFunction}
\eea
\end{widetext}

The appearance of $E+j\Omega$ in Eq.~\eqref{PhotonFullGaugeScalarGreenFunction} reflects the fact that the rotating frame couples to the angular momentum generator. For the auxiliary scalar sector, where $S_z=0$, the angular momentum entering this combination is simultaneously the orbital and total angular momentum. For the complete vector photon modes, by contrast, the same total angular momentum $j$ is shared by components with different orbital angular momenta. Thus, the single label $j$ appearing in the scalar Green function should not be interpreted as implying that every vector component has orbital angular momentum $j$.

The vector propagator can now be constructed using the scalar Green function. In inertial Minkowski coordinates, the Feynman-gauge operator reduces to the scalar d'Alembertian acting independently on each component of the vector field. Thus, the inertial-frame photon propagator has the tensor structure
\bea
D_{\rm I}^{\alpha\beta}(X,X')=\eta^{\alpha\beta}G_{\rm I}(X,X').
\label{PhotonFullGaugeInertialPropagator}
\eea
where $G_{\rm I}$ is the scalar Feynman Green function in inertial coordinates. Since the scalar Green function is a biscalar under coordinate transformations, its spacetime dependence is represented by $G(x,x')$ in the rotating coordinates.

Since the photon propagator is a rank-two bitensor, its transformation to the rotating coordinates involves a Jacobian at each spacetime point. We define
\bea
\Lambda^\mu{}_\alpha(x)\equiv\frac{\partial x^\mu}{\partial X^\alpha}.
\label{PhotonFullGaugeJacobian}
\eea
The corresponding transformation law is
\bea
D^{\mu\nu}(x,x')=\Lambda^\mu{}_\alpha(x)\Lambda^\nu{}_\beta(x')D_{\rm I}^{\alpha\beta}(X,X').
\label{PhotonFullGaugeBitensorTransformation}
\eea
and therefore
\bea
D^{\mu\nu}(x,x')=\Lambda^\mu{}_\alpha(x)\Lambda^\nu{}_\beta(x')\eta^{\alpha\beta}G(x,x').
\label{PhotonFullGaugeRotatingPropagator}
\eea

Substituting Eq.~\eqref{PhotonFullGaugeScalarGreenFunction} gives
\begin{widetext}
\bea
D^{\mu\nu}(x,x')=\frac{\ii}{(2\pi)^2}\Lambda^\mu{}_\alpha(x)\Lambda^\nu{}_\beta(x')\eta^{\alpha\beta}\sumint_j[dq]\int_{-\infty}^{+\infty}\frac{dE}{2\pi}\frac{e^{-\ii E(t-t')}e^{\ii j(\phi-\phi')}e^{\ii q_z(z-z')}J_j(|\mathbf q_\perp|\rho)J_j(|\mathbf q_\perp|\rho')}{(E+j\Omega)^2-|\mathbf q|^2+\ii\epsilon}.
\label{PhotonFullGaugeSpectralPropagator}
\eea
where the symbol $\sumint_j[dq]$ is defined in Eq.~\eqref{PhotonReducedModeMeasure}, and the tensor factor takes the explicit form
\bea
g^{\mu\nu}(x,x')\equiv\Lambda^\mu{}_\alpha(x)\Lambda^\nu{}_\beta(x')\eta^{\alpha\beta}
=
\begin{pmatrix}
1 & \Omega y' & -\Omega x' & 0\\
\Omega y & \Omega^2yy'-\cos\!\big[\Omega(t-t')\big] & -\Omega^2yx'-\sin\!\big[\Omega(t-t')\big] & 0\\
-\Omega x & -\Omega^2xy'+\sin\!\big[\Omega(t-t')\big] & \Omega^2xx'-\cos\!\big[\Omega(t-t')\big] & 0\\
0 & 0 & 0 & -1
\end{pmatrix}.
\label{PhotonFullGaugeBitensor}
\eea
\end{widetext}

We now verify that the propagator satisfies the Green-function equation of the gauge-fixed vector operator. In inertial coordinates,
\bea
\mathcal K_{\rm I}^\alpha{}_\gamma=\delta^\alpha_\gamma\Box_{\rm I},
\label{PhotonFullGaugeInertialOperator}
\eea
and
\bea
\mathcal K_{\rm I}^\alpha{}_\gamma(X)D_{\rm I}^{\gamma\beta}(X,X')=-\ii\eta^{\alpha\beta}\delta^{(4)}(X-X').
\label{PhotonFullGaugeInertialGreenEquation}
\eea

Under the coordinate transformation, covariance of the vector Green-function equation gives
{\small
\bea
\mathcal K^\mu{}_\lambda(x)D^{\lambda\nu}(x,x')=-\ii\Lambda^\mu{}_\alpha(x)\Lambda^\nu{}_\beta(x')\eta^{\alpha\beta}\delta^{(4)}(X-X').
\nn\\
\label{PhotonFullGaugeTransformedGreenEquation}
\eea
}

For the rotating transformation considered here, the Jacobian determinant has unit magnitude,
\bea
\left|\frac{\partial X}{\partial x}\right|=1,
\label{PhotonFullGaugeJacobianDeterminant}
\eea
and therefore
\bea
\delta^{(4)}(X-X')=\delta^{(4)}(x-x').
\label{PhotonFullGaugeDeltaTransformation}
\eea

Since the delta function has support only at $x=x'$, the two Jacobians can be evaluated at the same spacetime point. The metric transformation law then gives
\bea
\Lambda^\mu{}_\alpha(x)\Lambda^\nu{}_\beta(x)\eta^{\alpha\beta}=g^{\mu\nu}(x).
\label{PhotonFullGaugeMetricTransformation}
\eea
Hence,
\bea
\mathcal K^\mu{}_\lambda(x)D^{\lambda\nu}(x,x')=-\ii g^{\mu\nu}(x)\delta^{(4)}(x-x').
\label{PhotonFullGaugeGreenEquation}
\eea

In Feynman gauge, using $\mathcal K^\mu{}_\lambda=\delta^\mu_\lambda\nabla_\beta\nabla^\beta$, this equation can equivalently be written as
\bea
\nabla_\beta\nabla^\beta D^{\mu\nu}(x,x')=-\ii g^{\mu\nu}(x)\delta^{(4)}(x-x').
\label{PhotonFullGaugeCovariantGreenEquation}
\eea
where the covariant derivatives act on the vector index $\mu$ and therefore contain the Christoffel connections of the rotating coordinates. The scalar d'Alembertian $\Box$ enters only through the scalar Green function used in constructing the propagator.

It is worth emphasizing that the angular momentum label $j$ appearing in the spectral representation above labels the complete angular momentum sector. In the auxiliary scalar Green function, $j$ coincides with the orbital angular momentum because the scalar field has no spin contribution. When this scalar kernel is used to construct the vector propagator, the tensorial structure is supplied by the Jacobians $\Lambda^\mu{}_\alpha(x)$ and $\Lambda^\nu{}_\beta(x')$. Then, the appearance of a single Bessel order $J_j$ in the scalar kernel does not imply that the circular components of the corresponding vector modes all have orbital angular momentum $j$. Those components belong to the same total angular momentum sector $j$, but carry orbital angular momenta $j-1$ and $j+1$, respectively.

The full gauge-fixed propagator constructed here and the physical propagator obtained in Sec.~\ref{subsec:PhotonPropagator} serve different purposes. The physical propagator is constructed from the two independent physical photon modes and is therefore appropriate for correlation functions involving physical photon states and for processes in which only the physical transverse photon degrees of freedom are required. It is not, by itself, the Green function that inverts the complete gauge-fixed Maxwell operator and therefore should not be used as the internal photon propagator in a covariant perturbative formulation of QED.

In contrast, the full gauge-fixed propagator $D^{\mu\nu}$ is defined by the inverse of the complete gauge-fixed Maxwell operator and is therefore the appropriate Green function for perturbative calculations involving internal photon lines. Its tensor structure includes the gauge-fixed vector degrees of freedom required for the inversion of the operator and is not restricted to the two physical polarization modes. It should not, however, be interpreted as a propagator describing additional physical photon polarizations. The distinction is therefore between the physical-state propagator, which isolates the physical photon sector, and the gauge-fixed propagator, which provides the Green function required for the covariant formulation of the electromagnetic field.

\subsection{Alternative derivation from the Fock--Schwinger Representation of the Green Function}\label{secIXB}

Having established the complete set of vector eigenfunctions of Eqs.~\eqref{Psi+},~\eqref{Psi-},~\eqref{Psiz}, and~\eqref{Psi0} given by
\bea
\Psi_{a,j}^\mu(x)=e_a^\mu(x)u_j(x),
\eea
with
\bea
\nabla_\alpha\nabla^\alpha\Psi_{a,j}^\mu(x)=\lambda_j\Psi_{a,j}^\mu(x),
\eea
where
\bea
\lambda_j=-\left[(E+j\Omega)^2-|\mathbf q|^2\right],
\eea
we construct the Green function of the gauge-fixed Maxwell operator in Feynman gauge. The index $a$ labels the four covariantly constant vectors of the inertial Cartesian tetrad, while $j$ labels the angular dependence carried by the scalar cylindrical mode.

In Feynman gauge, the Green function $D^{\mu\nu}(x,x')$ is defined through
\bea
\nabla_\alpha\nabla^\alpha D^{\mu\nu}(x,x')=-\ii g^{\mu\nu}(x)\delta^{(4)}(x-x').
\eea
It is useful to introduce the vector differential operator
\bea
\mathcal{K}^\mu{}_\nu=\delta^\mu_\nu\nabla_\alpha\nabla^\alpha,
\eea
so that the Green equation takes the form
\bea
\mathcal{K}^\mu{}_\alpha D^{\alpha\nu}(x,x')=-\ii g^{\mu\nu}(x)\delta^{(4)}(x-x').
\eea
The vector eigenfunctions introduced above then satisfy
\bea
\mathcal{K}^\mu{}_\nu\Psi_{a,j}^\nu(x)=\lambda_j\Psi_{a,j}^\mu(x).
\eea
The scalar functions entering these eigenfunctions are
\bea
u_j(x)=J_j(|\mathbf{q}_\perp|\rho)e^{-\ii Et}e^{\ii j\phi}e^{\ii q_z z}.
\eea
For the purpose of the spectral construction, $E$, $|\mathbf{q}_\perp|$, and $q_z$ are taken as continuous spectral variables. The scalar cylindrical modes form a complete basis, with
\bea
\delta^{(4)}(x-x')=\sumint_j[dqdE]&u_j(x)u_j^*(x'),\nn\\
\eea
where
\bea
\sumint_j[dqdE]\equiv\sum_{j=-\infty}^{+\infty}\int_0^\infty\frac{|\mathbf{q}_\perp|d|\mathbf{q}_\perp|}{2\pi}\int_{-\infty}^{+\infty}\frac{dE}{2\pi}\int_{-\infty}^{+\infty}\frac{dq_z}{2\pi}\nn\\
\eea

The completeness of the corresponding vector basis follows from the tetrad decomposition of the inverse metric,
\bea
g^{\mu\nu}(x)=\eta^{ab}e_a^\mu(x)e_b^\nu(x).
\eea
Indeed, substituting $\Psi_{a,j}^\mu=e_a^\mu u_j$ into the spectral sum gives
 \bea
&&\sumint_j[dqdE]\eta^{ab}\Psi_{a,j}^\mu(x)\Psi_{b,j}^{\nu *}(x')\nn\\
&=&\eta^{ab}e_a^\mu(x)e_b^\nu(x')\delta^{(4)}(x-x').
\eea

Because the delta distribution has support only at $x'=x$, any smooth function multiplying it can be evaluated at $x'=x$. Therefore,
\bea
\eta^{ab}e_a^\mu(x)e_b^\nu(x')\delta^{(4)}(x-x')&=&\eta^{ab}e_a^\mu(x)e_b^\nu(x)\delta^{(4)}(x-x')\nn\\
&=&g^{\mu\nu}(x)\delta^{(4)}(x-x').
\eea
The contraction with $\eta^{ab}$ is essential here because the tetrad index $a$ labels the four components of the vector basis with the Minkowski metric $\eta^{ab}$, while the remaining indices $\mu$ and $\nu$ are spacetime coordinate indices. Since the tetrad is real in the rotating coordinates used here, complex conjugation acts only on the scalar mode, although the expression has been kept in its general form.

The inverse of the operator $\mathcal{K}$ can now be represented using the Fock--Schwinger proper-time method. For an eigenfunction with eigenvalue $\lambda_j$, the corresponding scalar factor in the Green function is
\bea
-\frac{\ii}{\lambda_j-\ii\epsilon},
\eea
where $\epsilon>0$ specifies the Feynman prescription. This factor admits the proper-time representation
\bea
-\frac{\ii}{\lambda_j-\ii\epsilon}=\int_0^\infty ds\,e^{-\ii s(\lambda_j-\ii\epsilon)}.
\eea
To verify this relation explicitly, we write
\bea
\int_0^\infty ds\,e^{-\ii s(\lambda_j-\ii\epsilon)}=\int_0^\infty ds\,e^{-\ii\lambda_js}e^{-\epsilon s}.
\eea
The factor $e^{-\epsilon s}$ ensures convergence as $s\rightarrow+\infty$. The integral can then be evaluated as
\bea
\int_0^\infty ds\,e^{-(\epsilon+\ii\lambda_j)s}=\frac{1}{\epsilon+\ii\lambda_j}=-\frac{\ii}{\lambda_j-\ii\epsilon}.
\eea
Thus, the Green function can be represented as the proper-time kernel
\bea
D^{\mu\nu}(x,x')=\int_0^\infty ds\,\left[e^{-\ii s(\mathcal{K}-\ii\epsilon)}\right]^{\mu\nu}(x,x').
\eea
The spectral decomposition of this kernel follows directly from the eigenvalue equation. In fact,
\bea
\mathcal{K}^\mu{}_\nu\Psi_{a,j}^\nu(x)=\lambda_j\Psi_{a,j}^\mu(x)
\eea
implies that the exponential operator acts diagonally on each eigenfunction,
\bea
e^{-\ii s(\mathcal{K}-\ii\epsilon)}\Psi_{a,j}^\mu(x)=e^{-\ii s(\lambda_j-\ii\epsilon)}\Psi_{a,j}^\mu(x).
\eea

The completeness relation therefore allows the kernel of the evolution operator to be expanded in the vector eigenbasis as
    \bea
&&\left[e^{-\ii s(\mathcal{K}-\ii\epsilon)}\right]^{\mu\nu}(x,x')\nn\\
&=&\sumint_j[dqdE]\eta^{ab}\Psi_{a,j}^\mu(x)\Psi_{b,j}^{\nu *}(x')e^{-\ii s(\lambda_j-\ii\epsilon)}.
\eea

Substitution into the proper-time representation after
performing the proper-time integral yields
\bea
D^{\mu\nu}(x,x')=-\ii\sumint_j[dqdE]\frac{\eta^{ab}\Psi_{a,j}^\mu(x)\Psi_{b,j}^{\nu *}(x')}{\lambda_j-\ii\epsilon}.
\eea
Using the explicit eigenvalue,
\bea
\lambda_j=-\left[(E+j\Omega)^2-|\mathbf q|^2\right],
\eea
the spectral denominator becomes
\bea
-\frac{\ii}{\lambda_j-\ii\epsilon}=\frac{\ii}{(E+j\Omega)^2-|\mathbf q|^2+\ii\epsilon}.
\eea
The Green function consequently takes the form
\bea
D^{\mu\nu}(x,x')=\ii\sumint_j[dqdE]\frac{\eta^{ab}\Psi_{a,j}^\mu(x)\Psi_{b,j}^{\nu *}(x')}{(E+j\Omega)^2-|\mathbf q|^2+\ii\epsilon}.
\eea
The numerator separates naturally into a tensorial contribution associated with the tetrad and a scalar contribution associated with the cylindrical modes,
\bea
\eta^{ab}\Psi_{a,j}^\mu(x)\Psi_{b,j}^{\nu *}(x')=\eta^{ab}e_a^\mu(x)e_b^\nu(x')u_j(x)u_j^*(x').
\eea

The scalar product appearing in the spectral representation is
\bea
u_j(x)u_j^*(x')&=&J_j(|\mathbf{q}_\perp|\rho)J_j(|\mathbf{q}_\perp|\rho')e^{-\ii E(t-t')}\nn\\
&\times&e^{\ii j(\phi-\phi')}e^{\ii q_z(z-z')}.
\eea
Combining these expressions, the Green function of Eq.~\eqref{PhotonFullGaugeSpectralPropagator} is recovered. 

The Green function can be verified directly by acting on its first spacetime argument with the vector covariant Laplacian. From the eigenvalue equation,
\bea
\nabla_\alpha\nabla^\alpha\Psi_{a,j}^\mu(x)=\lambda_j\Psi_{a,j}^\mu(x),
\eea
we obtain
\bea
\nabla_\alpha\nabla^\alpha D^{\mu\nu}(x,x')=-\ii\sumint_j[dqdE]\frac{\lambda_j\eta^{ab}\Psi_{a,j}^\mu(x)\Psi_{b,j}^{\nu *}(x')}{\lambda_j-\ii\epsilon}.\nn\\
\eea
For $\epsilon\rightarrow0^+$,
\bea
\frac{\lambda_j}{\lambda_j-\ii\epsilon}\longrightarrow1,
\eea
away from the distributional support associated with the pole prescription. The resulting expression is therefore
\bea
\nabla_\alpha\nabla^\alpha D^{\mu\nu}(x,x')=-\ii\sumint_j[dqdE]\eta^{ab}\Psi_{a,j}^\mu(x)\Psi_{b,j}^{\nu *}(x').\nn\\
\eea
Using the vector completeness relation established above,
\bea
\nabla_\alpha\nabla^\alpha D^{\mu\nu}(x,x')=-\ii g^{\mu\nu}(x)\delta^{(4)}(x-x'),
\eea
which reproduces the defining Green equation. The spectral representation is therefore consistent with the gauge-fixed Maxwell operator and its vector completeness relation.
}
\\

\subsection{The rigid rotation approximation}\label{secIXC}

We now consider a useful approximation of the scalar part of the full
gauge-fixed photon propagator in Eq.~\eqref{PhotonFullGaugeSpectralPropagator}.
We first perform the change of integration variable
$E\to E-l\Omega$. Since the energy integration extends over the entire real
axis, the integration measure remains unchanged,
\bea
dE\to dE,
\eea
whereas the denominator becomes
\bea
(E+l\Omega)^2-|\mathbf q|^2+\ii\epsilon\to E^2-|\mathbf q|^2+\ii\epsilon.
\eea
The temporal exponential transforms according to
\bea
e^{-\ii(E-l\Omega)(t-t')}=e^{-\ii E(t-t')}e^{\ii l\Omega(t-t')}.
\eea
Therefore, Eq.~\eqref{PhotonFullGaugeSpectralPropagator} can be written as
\bea
D^{\mu\nu}(x,x')=\frac{\ii}{(2\pi)^2}\Lambda^\mu{}_\alpha(x)\Lambda^\nu{}_\beta(x')\eta^{\alpha\beta}\widetilde D(x,x'),
\label{PhotonFullGaugeShiftedRepresentation}
\eea
where
\begin{widetext}
\bea
\widetilde D(x,x')=\sumint_l[dq]\int_{-\infty}^{+\infty}\frac{dE}{2\pi}\frac{e^{-\ii E(t-t')}e^{\ii l[\phi-\phi'+\Omega(t-t')]}e^{\ii q_z(z-z')}J_l(|\mathbf q_\perp|\rho)J_l(|\mathbf q_\perp|\rho')}{E^2-|\mathbf q|^2+\ii\epsilon}
\label{PhotonFullGaugeShiftedScalarRepresentation}
\eea
\end{widetext}

We now impose the rigid-rotation condition, according to which the relative
azimuthal displacement of the two points is determined by the angular
velocity and the elapsed time,
\bea
\Omega(t-t')+(\phi-\phi')=0.
\label{PhotonRigidRotationCondition}
\eea

This condition corresponds to the rigid-rotation approximation, in which the relative azimuthal displacement of the two points is determined by the angular velocity over the time interval considered. A similar approximation has been employed in Refs.~\cite{Ayala:2026wro,CastanoYepes:2025} to describe fermion propagation in a rigidly rotating cylindrical environment. Its applicability requires that radial expansion can be neglected, so that the evolution of the system can be approximated by rigid rotation over the time interval considered. In the context of relativistic heavy-ion collisions, this approximation is relevant during stages of the evolution in which radial expansion remains subleading.

Under this condition, the angular-momentum-dependent phase in
Eq.~\eqref{PhotonFullGaugeShiftedScalarRepresentation} becomes the unity and the remaining sum over the cylindrical angular momentum can be evaluated
using the addition theorem for Bessel functions,
\bea
\sum_{l=-\infty}^{\infty}J_l(|\mathbf q_\perp|\rho)J_l(|\mathbf q_\perp|\rho')=J_0\left(|\mathbf q_\perp|r\right),
\label{PhotonBesselAdditionTheorem}
\eea
where the radial separation is defined by
\bea
r\equiv|\rho-\rho'|.
\label{PhotonRadialSeparation}
\eea

This zeroth-order Bessel function can be represented as an angular integral,
\bea
J_0(|\mathbf q_\perp|r)=\frac{1}{2\pi}\int_0^{2\pi}d\varphi_q\,e^{-\ii|\mathbf q_\perp|r\cos\varphi_q},
\label{PhotonBesselAngularRepresentation}
\eea
which is the angular representation associated with the Jacobi--Anger
expansion. Substituting Eq.~\eqref{PhotonBesselAngularRepresentation} into
Eq.~\eqref{PhotonFullGaugeShiftedScalarRepresentation} gives
\begin{widetext}
\bea
\widetilde D(x,x')\to\frac{1}{2\pi}\int_{-\infty}^{\infty}dq_z\int_0^\infty|\mathbf q_\perp|\,d|\mathbf q_\perp|\int_0^{2\pi}d\varphi_q\int_{-\infty}^{+\infty}\frac{dE}{2\pi}\frac{e^{-\ii E(t-t')}e^{\ii q_z(z-z')}e^{-\ii|\mathbf q_\perp|r\cos\varphi_q}}{E^2-|\mathbf q|^2+\ii\epsilon}.
\label{PhotonRigidRotationAngularFourierRepresentation}
\eea
\end{widetext}

The transverse momentum measure is the usual cylindrical measure,
\bea
d^2\mathbf q_\perp=|\mathbf q_\perp|\,d|\mathbf q_\perp|\,d\varphi_q,
\label{PhotonTransverseMomentumMeasure}
\eea
so that the preceding expression can be written as
\bea
&&\widetilde D(x,x')\nn\\
&\to&\frac{1}{2\pi}\int d^3q\int_{-\infty}^{+\infty}\frac{dE}{2\pi}\frac{e^{-\ii E(t-t')}e^{\ii q_z(z-z')}e^{-\ii|\mathbf q_\perp|r\cos\varphi_q}}{E^2-|\mathbf q|^2+\ii\epsilon}.\nn\\
\label{PhotonRigidRotationFourierRepresentation}
\eea

To make the Fourier structure explicit, consider the corresponding orthogonality relation for the cylindrical Bessel functions Eq.~\eqref{PhotonFullGaugeCylindricalCompleteness} together with
\bea
\int_0^{2\pi}d\varphi_q\,e^{-\ii|\mathbf q_\perp|r\cos\varphi_q}=2\pi J_0(|\mathbf q_\perp|r),
\label{PhotonAngularBesselIdentity}
\eea
this establishes the correspondence between the cylindrical representation
and the transverse Fourier transform. In particular, the angular integral
reconstructs the transverse plane-wave factor, while the Bessel orthogonality
relation reconstructs the transverse momentum measure.

Introducing the four-momentum
\bea
q^\mu=(E,\mathbf q),
\label{PhotonFourMomentum}
\eea
the resulting scalar propagator has the Fourier representation
\bea
\widetilde D(x-x')=(2\pi)^2\int d^4q\,\frac{e^{-\ii q\cdot(x-x')}}{q^2+\ii\epsilon},
\label{PhotonFullGaugeFourierScalarPropagator}
\eea

Therefore, after the rigid-rotation approximation, the scalar part acquires the
standard translationally invariant Fourier structure. The dependence on the
individual coordinates is reduced to their relative separation.

The full gauge-fixed photon propagator can therefore be written in the
factorized form
\bea
D^{\mu\nu}(x,x')=\frac{\ii}{(2\pi)^2}g^{\mu\nu}(x,x')\widetilde D(x-x'),
\label{finalphotonpropapprox}
\eea

Its explicit expression for the tensor structure is given in Eq.~\eqref{PhotonFullGaugeBitensor}. To
evaluate this bitensor under the rigid-rotation condition, we use $\phi-\phi'=-\Omega(t-t')$ together with the cylindrical representation of the transverse coordinates,
\bea
x&=&\rho\cos\phi,\nn\\
x'&=&\rho'\cos\phi',\nn\\
y&=&\rho\sin\phi,\nn\\
y'&=&\rho'\sin\phi'.
\label{PhotonTransverseCartesianCoordinates}
\eea
The propagator is therefore expressed as the product of a scalar Fourier
kernel, which has the standard translationally invariant momentum dependence,
and a coordinate-dependent bitensor generated by the transformation from
inertial to rotating coordinates. The latter retains the dependence on the
rotating coordinate system, while the scalar factor has the same momentum-space
structure as the free Minkowski-space propagator.

\section{Summary and conclusions}\label{sec:conclusiones}

We have constructed the quantized electromagnetic field in a rigidly rotating reference frame and derived the corresponding physical photon propagator. The transformation to rotating coordinates preserves the flatness of Minkowski spacetime, while the resulting metric incorporates the inertial effects associated with the rotating frame. The light cylinder defines the spatial region in which the rigidly rotating description can be consistently applied.

{In the present construction, the transverse mode expansion is formulated on the unbounded coordinate domain, so that the radial momentum remains continuous and no boundary condition at a finite cylindrical radius is imposed. Accordingly, the propagators obtained here should be understood as rotating-coordinate Green functions applicable locally within the causal region $\Omega\rho<1$, rather than as normal-mode Green functions for an electromagnetic field confined to a finite rotating cylinder. The latter problem would require specifying boundary conditions at the cylinder surface and would in general lead to a discrete transverse spectrum.}

The covariant Maxwell equations were formulated in the rotating geometry with a covariant gauge-fixing term. In Feynman gauge, the field equations reduce to the covariant vector wave equation. The rotational properties of the vector field require distinguishing the orbital and intrinsic contributions to angular momentum. Introducing the transverse rotation generator makes this structure explicit and identifies the total angular momentum as the appropriate quantum number for the vector modes. The two circular transverse components then carry different orbital angular momenta while belonging to the same total angular momentum sector.

The physical solutions were obtained in cylindrical coordinates in terms of regular Bessel functions. After imposing the physical conditions on the gauge-fixed field, two independent photon modes remain for each angular momentum sector. Their explicit form shows how the circular polarization components are combined with different orbital contributions, while the second mode also contains a longitudinal spatial component. This construction provides a basis of physical solutions adapted to the rotational symmetry of the system.

The relation between the energy measured with respect to the rotating time coordinate and the inertial frequency was established through the angular-momentum-dependent rotational shift. This distinction enters directly into the mode normalization, which was obtained using the Maxwell inner product. The two physical polarization sectors are orthogonal and can therefore be quantized independently. Promoting their mode coefficients to bosonic operators leads to the canonical commutation relations and defines the corresponding photon vacuum and Fock states.

The physical propagator was constructed directly from the quantized modes. The greater and lesser two-point functions are expressed as sums over the angular momentum sectors and integrations over the cylindrical momentum variables. The transverse dependence of each polarization sector can be collected into polarization kernels, whose sum contains the complete physical polarization structure of the propagating field. This representation keeps the radial dependence and angular momentum content explicit, rather than absorbing them into a translationally invariant momentum-space tensor.

The time-ordered physical propagator was then written in spectral form. Its pole structure contains the rotational energy shift associated with the angular momentum of each mode, while the two time-ordering contributions retain the appropriate opposite pole prescriptions. The resulting expression provides a spectral representation in which the temporal dependence, longitudinal propagation, and transverse polarization structure are separated into distinct factors. We also constructed the full gauge-fixed photon propagator by first obtaining the scalar Green function in the rotating coordinates and then using the tensor transformation of the inertial Feynman-gauge propagator. The physical propagator is obtained from the two independent physical photon modes and is therefore suited to correlation functions and processes involving physical photon states, with their polarization and angular momentum structure explicitly resolved. The full gauge-fixed propagator, in contrast, is the Green function that inverts the complete gauge-fixed Maxwell operator and is the appropriate object for covariant perturbative calculations with internal photon lines. We verified explicitly that it satisfies the corresponding covariant Green-function equation in the rotating coordinates. The two propagators thus describe different aspects of the electromagnetic field, with the former restricted to the physical photon sector and the latter providing the gauge-fixed Green function required by the field-theoretic formulation. Under the rigid rotation approximation, the fixed-gauge photon propagator can be expressed as the product of the usual Minkowski space translationally invariant scalar piece times the space-dependent metric tensor which contains the effects of the loss of Lorentz invariance in the rotating environment.

The formalism developed here provides a framework for studying electromagnetic propagation and correlation functions in rigidly rotating systems while retaining the angular momentum structure of the photon modes explicitly. It can be extended to situations in which the electromagnetic field interacts with rotating matter, where the distinction between inertial and rotating-frame energies and the spatial structure of the photon modes can enter the resulting observables.

\section*{Acknowledgments}

JDC-Y acknowledges the kind support received from L.A. Hern\'andez and the Universidad Aut\'onoma Metropolitana, as well as from the Instituto de Ciencias Nucleares (ICN), Universidad Nacional Aut\'onoma de M\'exico, during a work visit when the final stages of this work were carried out. He also gratefully acknowledges the Universidad del Valle for authorizing the medium-term research leave within the framework of PPI No.~71433 of the Vicerrector\'ia de Investigaciones, which made this work visit to the ICN possible. He further thanks the Isaac Newton Institute for Mathematical Sciences, University of Cambridge, for support and hospitality during the programme \textit{Quantum field theory with boundaries, impurities, and defects (BID2025)}, which provided the time and working environment in which the initial calculations leading to this work were undertaken. S.B.-L. and J.J.M.-S. each acknowledge the financial support of fellowships granted by Secretar\'ia de Ciencia, Humanidades, Tecnolog\'ia e Innovaci\'on (SECIHTI) M\'exico as part of the Sistema Nacional de Posgrados. Support for this work has also been received in part by DGAPA-PAPIIT-UNAM grant number AG100826, and from Secretar\'ia de Ciencia, Humanidades, Tecnolog\'ia e Innovaci\'on (SECIHTI) M\'exico grant numbers CIORGANISMOS-2025-17, CBF-2025-G-1718, and CBF-2026-465.

\bibliography{biblioROT}

\end{document}